\documentclass[amsmath,amssymb,floatfix,aps,superscriptaddress,twocolumn,nofootinbib,notitlepage,prx]{revtex4-2}

\usepackage{amsmath,amsthm,amssymb,bm,graphicx,float,xcolor,changes,mathtools,enumitem,booktabs}
\usepackage[colorlinks=true,urlcolor=purple,linkcolor=blue,citecolor=green,bookmarks=false]{hyperref}
\allowdisplaybreaks

\newcommand{\ra}[1]{\renewcommand{\arraystretch}{#1}}

\newcommand{\be}{\begin{equation}}
\newcommand{\ee}{\end{equation}}
\newcommand{\6}{\partial}

\usepackage{pdfpages} 
\makeatletter
\AtBeginDocument{\let\LS@rot\@undefined}
\makeatother

\begin{document}

\title{Numerical methods and analytic results for one-dimensional strongly interacting spinor gases}

\author{Ovidiu I. P\^{a}\c{t}u}
\affiliation{Institute for Space Sciences, Bucharest-M\u{a}gurele, R 077125, Romania}

\begin{abstract}

One of quantum physics' fundamental, but largely unsolved, problems is the computation of the correlation functions in many-body systems.
In this paper we address this problem in the case of one-dimensional spinor gases with repulsive contact interactions in the presence of a 
confining potential. We take advantage of the fact that in the strong coupling limit, the wavefunction factorizes with the charge degrees of 
freedom expressed as a Slater determinant of spinless fermions and the spin sector described by a spin chain of Sutherland type with exchange
coefficients that depend only on the trapping potential. This factorization is also present in the expressions for the correlation functions. 
Still, analytical and numerical investigations were hindered by the fact that the local exchange coefficients and the charge component of the 
correlators are expressed as $N-1$ multidimensional integrals, with $N$ the number of particles, which are notoriously hard to compute using 
conventional methods. We introduce a new approach to calculating these integrals that is extremely simple, scales polynomially with the number 
of particles, and is several orders of magnitude faster than the previous methods reported in the literature. This allows us to investigate 
the static and dynamic properties, temperature dependence, and nonequilibrium dynamics for systems with a larger number of particles than 
previously considered and discover new phenomena. We show that, contrary to natural expectations, the momentum distribution of strongly 
interacting trapped spinor gases becomes narrower as we increase the temperature and derive simple determinant representations for the 
correlators in the spin incoherent regime valid for both equilibrium and nonequilibrium situations.  For homogeneous systems of impenetrable 
particles at zero temperature we analytically compute the large distance asymptotics of the correlators finding that the leading term of the 
asymptotic expansion is proportional to $e^{-x k_F\ln \kappa/\pi}$,  with $\kappa$  the number of components of the system, which precludes any 
singularity of the momentum distribution.

\end{abstract}

\maketitle

\section{Introduction}

One-dimensional (1D) quantum systems have received considerable interest in the last decades from both theoretical and experimental communities 
\cite{BDZ08,GBL13,MVBF23}. From the theoretical point of view this interest stems from the observation that quantum effects are enhanced in low 
dimensions coupled with the fact that many 1D quantum systems are integrable opening the way for comprehensive analytical and numerical 
investigations of the correlation functions and other physical quantities of interest \cite{KBI,EFGKK}. On the experimental side the 
unprecedented degree of control achieved in the manipulation of utracold atomic gases resulted in the experimental realization of many 
paradigmatical 1D models \cite{PWMM04,KWW04,PMCL14}  followed by  subsequent investigations of many new intriguing  phenomena specific to 
low-dimensional systems.  

1D multicomponent systems are of particular importance because they present  phases not found in their higher-dimensional counterparts and 
exhibit exotic phenomena like spin-charge separation \cite{G03}. In addition, their dynamics is expected to be very rich as a result of the 
interplay between the internal degrees of freedom, interaction, external fields and statistics of their constituents. In this paper we will 
consider spinor gases, bosonic or fermionic, with repulsive contact interactions subjected to a confining potential. These systems, also known as 
$SU(\kappa)$ gases, with $\kappa$ the number of components (internal degrees of freedom), are routinely  realized in laboratories in both the 
few-body \cite{ZSLW12} and many-body \cite{PMCL14} limits. In the homogeneous case (no trapping potential) the spinor gases are integrable and 
in particular cases they reduce to some of the most well known models of 1D physics: the Lieb-Liniger model \cite{LL63} for $\kappa=1$, the 
Gaudin-Yang model \cite{G67,Y67} for $\kappa=2$, and the Sutherland models \cite{S68} for $\kappa\ge 3$. Fermionic systems have an 
antiferromagnetic ground state in  accordance to  Lieb-Mattis theorem \cite{LM62} and at zero and low temperatures their correlation functions 
are described by bosonization  and Luttinger liquid (LL) theory \cite{H81,G03}. While Lieb-Liniger (spinless)  bosons are described at low 
temperatures by LL theory multicomponent bosons with spin independent interactions have a polarized ground state \cite{EL02}. In this case the 
softest excitation above the ground state is the magnon mode with a quadratic dispersion  relation which rules out the application of LL theory. 
This phase is a new low energy universality class called  ferromagnetic liquid \cite{ZCG07,AT07,MF08,KG09} with very distinct properties which 
are neither those of a localized ferromagnet nor of a Luttinger liquid. In the case of strong interactions the energy of the spin sector becomes 
much smaller than the energy of the charge sector $E_{spin}\ll E_{charge}$. When the thermal energy is much  larger than the energy scale of the 
spin excitations but much smaller than the energy of the charge sector, $E_{spin}\ll k_BT\ll E_{charge}$, the fermionic and bosonic systems with 
$\kappa\ge 2$ are described by the spin incoherent Luttinger liquid universality class (SILL) \cite{BL,B1,CZ1,CZ2,Matv,FB04,Fiet07} which is has 
a higher degree of universality than the LL and distinct properties.

In the strong coupling limit the Bethe ansatz wavefunctions of a homogeneous system factorize in spin and charge components with the charge 
degrees of freedom described by  Slater determinants of free  fermions and the spin sector characterized by a Sutherland type spin chain. This 
factorization was first noticed by Ogata and Shiba \cite{OS90} in their investigation of the Hubbard model and represented the starting point of 
Izergin and Pronko's  derivation of determinant  representations for the correlators of the  Gaudin-Yang model \cite{IP98} in the SILL regime. 
Usually experiments are performed in the presence of a confining potential which breaks integrability. From general considerations it can be 
argued that the  high degeneracy of the groundstate and the factorization of the wavefunctions  remain valid for inhomogeneous systems with the 
charge degrees of freedom  now being  described by  Slater determinants of  spinless fermions  subjected to the same trapping \cite{DFBB08,GCWM09}. 
In the case of the spin sector it  took some  time until it was realized that the Sutherland like spin chain that characterizes the spin degrees 
of freedom should have position variable coefficients which reflect the inhomogeneity of the trapping potential  \cite{DBBR14,VFJV14,VPVF15,
LMBP15, YGP15,YP16,YC16}. 

There are several advantages to this factorized description of strongly interacting spinor gases. The effective spin chain  provides   insights 
into the magnetic properties of spinor gases and the reduction  of the Hilbert space allows for the investigation of  larger number of 
particles than the conventional diagonalization methods employed for continuum systems. The spin sector can be studied using exact diagonalization 
for lattice systems or the Density Matrix Renormalization Group while the charge sector being described by spinless fermions in principle should 
be efficiently computed. In addition, one can also investigate more efficiently the low temperature correlators and the nonequilibrium dynamics. 
There are, however, some caveats. Knowledge of the wavefunction, while important, does not mean that the relevant physical quantities like the 
densities or momentum distributions are easily accessible.  After all, we have known the wavefunctions of integrable systems like the Lieb-Liniger 
\cite{LL63} and Gaudin-Yang \cite{G67,Y67} models for almost 6o years and our understanding of their correlation functions is rather incomplete 
even in the case of single component systems while it is almost nonexistent for multicomponent models.   
The important feature is that the factorized nature of the wavefunction translates into the expressions for the correlation functions and one can 
compute the spin and charge components independently. One would expect that the computation of the spin functions would represent the limiting 
factor in the number of particles that can be investigated but this is not true. While the calculation of the spin sector properties is difficult 
at the present time the main computational bottleneck is represented by the calculation of the charge functions. The reason is that the position 
dependent coefficients appearing in the description of the effective spin chain, also known as local exchange coefficients, and the charge 
components of the corelators are expressed in terms of $N-1$  multidimensional integrals over products of Slater determinants, with $N$  the number 
of particles, which are very hard to compute.  Initial approaches used brute force or Monte-Carlo integration for systems  up to 16 particles 
\cite{DBBR14,VFJV14,VPVF15,LMBP15, YGP15,YP16,YC16,JY16,JY17} followed by the introduction of  more sophisticated  methods which allowed for the 
computation of the coefficients for systems up to  $N=35$ \cite{LKTZ16,LKTV16} and $N=60$ \cite{DBS16} particles. The main drawbacks of these 
methods are the need for arbitrary precision calculations and the fact that they are very time consuming. For example, the calculation of the 
coefficients for  $N=30$ takes about an hour \cite{LKTV16,DBS16} and more than a week for $N=60$ \cite{DBS16}.  These shortcomings are exacerbated 
in the case of nonequilibrium dynamics which requires the calculation of the local exchange coefficients at every step of the iterative process 
required to solve for the dynamics \cite{BFZ19,CTAM24}. The calculation of the charge functions for the correlators, also known as the one-body 
density matrix elements, is even more involved than  the case of the local exchange coefficients with results for up to 20 particles being 
reported in \cite{DBS16}. An elegant method of computing the one-body density matrix elements using the connection with the  correlation functions 
of 1D impenetrable anyons in the same geometry was introduced by Yang and Pu in \cite{YP17} allowing for the investigation of harmonically trapped 
systems with up to 60 particles (see also \cite{JY18}).

One of the main results of this paper is the introduction of an extremely efficient numerical method of calculating all the relevant charge 
functions: local exchange coefficients, single particle densities and one-body density matrix elements for any confining potential. Our method has 
a polynomial complexity in the number of particles, significantly outperforms all the other previously known algorithms, and is extremely simple 
requiring only the  calculation of determinants involving partial overlaps of the single particle basis and the Discrete Fourier Transform. It is 
also exact with the only source of errors being the accuracy with which the partial overlaps of the single particle orbitals can be computed and 
does not require the use of arbitrary precision subroutines. For example, we are able to compute  the local exchange coefficients in the case of 
harmonic trapping in less than 2 seconds for $N=60$ and in less than 30 seconds for $N=120$. The one-body  density matrix elements for 20 particles 
are computed in 0.015 seconds (compared with 742 seconds using the code provided in the ancillary files of \cite{DBS16}) and in less than  a second 
for $N=60$. 
We use this method to determine the density profiles and momentum distributions for harmonically trapped two-component fermionic and bosonic gases 
with  $N=26$ particles for different values of the spin imbalance. We extract the  Tan contacts of each component from the $ C_\sigma/k^4$ tails of 
the momentum distributions (the spectrum of the effective spin chain can be used to compute only the total contact) and 
show that for fermionic gases  the contacts decrease as a function of the spin imbalance with the maximum obtained for the balanced system while in 
the  case of bosonic gases the total contact is constant with  each individual contact increasing or decreasing linearly as a function of imbalance.

An important and interesting  feature of strongly interacting spinor gases in 1D is that small changes in temperature can produce dramatic changes 
in  the momentum  distributions, spectral functions and transport properties of such systems \cite{Fiet07}. We show that the transition from the 
LL/ferromagnetic 
liquid regime to the SILL phase is marked by a significant reconstruction of the momentum distribution in which the number of particles at large 
values of momenta decreases as the temperature is increased. This counterintuitive behavior, was probed indirectly in the case of two-component systems  
\cite{CSZ05,PKF18,PV20} but here we map the entire transition for two-, three-, and four-component systems and prove that is a general feature of 
multicomponent systems. The amplitude of this reconstruction is expected to decrease with the number of components. 
The way temperature influences the  transport properties is investigated by studying the nonequilibrium dynamics after a quench from a domain wall 
boundary state \cite{PVM22}. At zero temperature, after the quench, the integrated magnetization as a function  of time exhibits superdiffusive 
behaviour. At very small temperatures, for which the charge degrees of freedom are still close to the groundstate but the spin sector becomes 
incoherent, the integrated magnetization after the quench shows ballistic behaviour.  This highlights the oversized role of the temperature in 
influencing  the transport properties of strongly interacting spinor gases.

The main idea of our, mainly numerical, method can also be used to derive very efficient determinant representations for the space-, magnetic field-, 
and temperature-dependent correlators of impenetrable particles in the SILL regime. These representations are valid in both equilibrium and 
nonequilibrium situations generalizing the results of Pezer and Buljan \cite{PB07} at zero temperature and  Atas \textit{et al}, \cite{AGBK17} at 
finite temperature for the single component bosonic Tonks-Girardeau gas. In the case of  homogeneous systems at zero temperature we provide a Fredholm 
determinant representation in terms of the sine-kernel which contains as particular cases the well known results for single component bosons 
\cite{Schu63,L66} and two-component  fermions and bosons \cite{IP98}. We calculate analytically the large distance asymptotics of these correlators 
from the solution of the associated Riemann-Hilbert problem. The leading term  of the asymptotics is given by $e^{-x k_F\ln \kappa/\pi}$ implying that 
the momentum distributions of systems in the SILL regime do not present  singularities even at zero temperature. 

The plan of the paper is as follows. In Sec.~\ref{s2} we review the 1D spinor gas and its eigenstates in the strongly interacting regime. In 
Sec.~\ref{s3} we introduce the new method of computing the relevant charge quantities and in Sec.~\ref{s4} we present numerical results obtained 
using  this new method. The determinant representations for the trapped and homogeneous systems in the SILL regime are derived in Sec.~\ref{s5} and  
Sec.~\ref{s6}, respectively. We conclude in Sec.~\ref{s7}. Several technical details are relegated to four appendices.

\section{One-dimensional spinor gases}\label{s2}

We are interested in studying the static and dynamic properties of one-dimensional spinor gases with $\kappa$ components and strong repulsive contact  
interactions. In the presence of an external confining potential $V(z)$ the Hamiltonian in second quantization is
\begin{align}\label{ham}
H=&\int dz\,\frac{\hbar^2}{2m}\left(\6_z\boldsymbol{\Psi}^\dagger\6_z\boldsymbol{\Psi}\right)+\frac{g}{2}:\left(\boldsymbol{\Psi}^\dagger
\boldsymbol{\Psi}\right)^2:\nonumber\\
&\qquad \qquad \qquad +V(z)\left(\boldsymbol{\Psi}^\dagger\boldsymbol{\Psi}\right)-\boldsymbol{\Psi}^\dagger\boldsymbol{\mu}\boldsymbol{\Psi}\, ,
\end{align}
where $m$ is the mass of the particles,  $\boldsymbol{\Psi}=\left(\Psi_1(z),\cdots,\Psi_\kappa(z)\right)^T$, $\boldsymbol{\Psi}^\dagger=\left(
\Psi_1^\dagger(z),\cdots, \Psi_\kappa^\dagger(z)\right)$,  $:\ \  :$ represents normal ordering, $g>0$ is the strength of the repulsive contact 
interaction and $\boldsymbol{\mu}$ is a matrix which has the chemical potentials $\mu_1,\cdots,\mu_\kappa$  of each component  on the diagonal and 
zero otherwise.  We consider fermionic or bosonic gases with the field operators in (\ref{ham}) satisfying the commutation relations
\begin{subequations}
\begin{align}
\Psi_\sigma(x)\Psi_{\sigma'}^\dagger(y)-\epsilon\Psi_{\sigma'}^\dagger(y)\Psi_\sigma(x)&=\delta_{\sigma,\sigma'}\delta(x-y)\, ,
\label{com1}\\
\Psi_\sigma^\dagger(x)\Psi_{\sigma'}^\dagger(y)-\epsilon\Psi_{\sigma'}^\dagger(y)\Psi_\sigma^\dagger(x)&=0\, ,\label{com2}\\
\Psi_\sigma(x)\Psi_{\sigma'}(y)-\epsilon\Psi_{\sigma'}(y)\Psi_\sigma(x)&=0\, ,\label{com3}
\end{align}
\end{subequations}
where $\sigma,\sigma'\in\{1,\cdots,\kappa\}$ and $\epsilon=1\, (-1)$ in the bosonic (fermionic) case. We will mainly use as an example the case of 
harmonic confinement $V(z)=m\omega^2 z^2/2$ with $\omega$ the trapping frequency but almost all the results derived in the paper remain true for any 
well behaved trapping potential including the case of Dirichlet boundary conditions in a system of dimension $L$ which can be implemented by 
$V(z)=0$ for $z\in[-L/2,L/2]$ and $V(z)=\infty$ for $z\notin[-L/2,L/2]$. When the trapping potential is absent the Hamiltonian (\ref{ham}) is 
integrable for any value of the interaction strength and the wavefunctions and energy spectrum were obtained by Lieb and Liniger  in \cite{LL63} for 
the single component case, by Gaudin and Yang \cite{Y67,G67} for $\kappa=2$ and in the general case by Sutherland in \cite{S68}. When an external 
potential is present  the system is solvable only for $g=0$ and $g=\infty$. 

In some applications it will also be useful to add to the Hamiltonian (\ref{ham}) a spin dependent magnetic gradient term of the form  $-G z 
\boldsymbol{\Psi}^\dagger\sigma^z\boldsymbol{\Psi}$ \cite{DBBR14,YGP15,YP16} where $\sigma^z$ is a generalized Pauli matrix with elements $[\sigma^z]_{a,b}=
\hbar(\frac{\kappa-1}{2}+1-b)\delta_{a,b}$ and $G$ quantifies the  strength of the gradient. In first quantization the Hamiltonian of the spinor gas 
with the magnetic gradient term is
\begin{align}
\mathcal{H}=&\left[\sum_{j=1}^N-\frac{\hbar^2}{2m}\frac{\6^2}{\6z_j}+V(z_j)-G z_j\sigma^z_j\right]\nonumber\\
&\qquad\qquad+g \sum_{i<j}\delta(z_i-z_j)-\sum_{\sigma=1}^\kappa\mu_\sigma N_\sigma\, ,
\end{align}
where $N$ is the total number of particles and $N_\sigma$ is the number of particles of type $\sigma$. From now on  we will consider $\hbar=m=\omega
=1$.

\subsection{Groundstate manifold of eigenstates in the strongly interacting regime}

A general eigenstate of the Hamiltonian (\ref{ham}) is given by 
\begin{align}\label{eigenstate}
|\Phi\rangle&=\int dz_1 \cdots dz_N\sum_{\sigma_1,\cdots,\sigma_N=\{1,\cdots,\kappa\} }^{[N_1,\cdots, 
N_\kappa]} \psi^{\sigma_1\cdots\sigma_N}(z_1,\cdots,z_N)\nonumber\\
&\qquad\qquad\qquad\qquad\times\Psi_{\sigma_N}^\dagger(z_N)\cdots\Psi_{\sigma_1}^\dagger(z_1)|0\rangle\, ,
\end{align}
with $|0\rangle$  the  Fock vacuum satisfying $\Psi_\sigma(z)|0\rangle=0$ for all values of $\sigma$ and $z$. We will call such a state to be in the 
$\boldsymbol{N}$ sector where we have introduced the notation $\boldsymbol{N}=[N_1,\cdots, N_\kappa]$ with $\sum_{\sigma=1}^\kappa N_\sigma=N$. In 
Eq.~(\ref{eigenstate}) the  $[N_1,\cdots, N_\kappa]$  over the sum sign means that the sum is restricted such that the number of creation operators of 
type $\sigma$ is $N_\sigma$. For arbitrary repulsion an explicit expression for the wavefunction $\psi^{\sigma_1\cdots\sigma_N}(z_1,\cdots,z_N)$ is 
outside the reach of analytical methods but in the strongly interacting regime described by $g$ very large but finite, the charge and spin degrees of 
freedom factorize \cite{OS90,IP98,DFBB08,GCWM09,DBBR14,VFJV14,VPVF15,LMBP15,YAP22,MV22,OPCA23} and the wavefunction of the \textit{lowest lying  states} takes 
the form [$\boldsymbol{z}=(z_1,\cdots,z_N)$]
\begin{align}\label{wavef}
\psi^{\sigma_1\cdots\sigma_N}&(\boldsymbol{z})=\left[\sum_{P \in S_N} (-\epsilon)^P P\theta(\boldsymbol{z})
P\chi
(\sigma_1,\cdots,\sigma_{N})\right]\nonumber\\
&\qquad\qquad\qquad\qquad\qquad\times\psi_F(\boldsymbol{z}|\boldsymbol{q}^0)\, ,
\end{align}
with $\boldsymbol{q}^0\equiv(q^0_1,\cdots,q^0_N)=(1,\cdots,N)$ and
\be\label{slater}
\psi_F(\boldsymbol{z}|\boldsymbol{q}^0)=\frac{1}{\sqrt{N!}}\det_N\left[\phi_{q^0_j}(z_i)\right]_{i,j=1,\cdots,N}\, ,
\ee
is the Slater determinant formed from the lowest $N$ orbitals $\phi_{q^0_j}(z)$ of a system of spinless fermions subjected to  the same potential  $V(z)$ 
and  $\chi(\sigma_1,\cdots,\sigma_N)$ is an eigenfunction of a spin chain characterizing the spin degrees of freedom  which  we will  describe below. In 
Eq.~(\ref{wavef}) $\theta(\boldsymbol{z})\equiv\theta(z_1<\cdots< z_N)$ is a generalized Heaviside function which is 1  when $z_1<\cdots<z_N$  and zero  
otherwise and  $S_N$ is the group of permutations of $N$ elements with $(-1)^P$ the signature of the  permutation. For a given permutation  
$P=\left(\begin{array}{ccc} 1 &  \cdots & N\\
P_1&\cdots& P_N
\end{array}\right) $
the action of this permutation on $\theta(\boldsymbol{z})$ is given by $P\theta(\boldsymbol{z})=\theta(z_{P_1}<\cdots<z_{P_N})$ and $P\chi(\sigma_1,
\cdots, \sigma_N)=\chi(\sigma_{P_1}, \cdots, \sigma_{P_N})$. The spin chain whose wavefunctions describe the spin sector is (we consider the general case 
with a  gradient term)  \cite{DBBR14,VFJV14,VPVF15,LMBP15,YGP15,YP16,YC16}.
\be\label{sc}
H_{sc}^0=E_F(\boldsymbol{q}^0)-\frac{1}{g}\sum_{i=1}^{N-1} J_i^0\left(1+\epsilon \hat{P}_{i,i+1}\right)-G\sum_{i=1}^N D_i^0\sigma_i^z\, ,
\ee
where $E_F(\boldsymbol{q}^0)=\sum_{j=1}^N\varepsilon(q_j^0)$ is the groundstate energy of the spinless fermionic system, $\hat{P}_{i,i+1}$ is the  
operator  that permutes the spins on positions $i$ and $i+1$ and $J_i^0$ are the local exchange coefficients  which can be obtained from the  Slater 
determinant 
$\psi_F(\boldsymbol{q}^0)$ via \cite{VFJV14}
\begin{align}\label{coeff} 
J_i^0=&N!\int dz_1\cdots dz_N\,\delta(z_i-z_{i+1})\,\theta(z_1<\cdots<z_N)\nonumber\\
&\qquad\qquad \times \left|\frac{\6 \psi_F(\boldsymbol{q}^0)}{\6 z_i}\right|^2\, , \
 i=1,\cdots,N-1\, . 
\end{align}
The coefficients $D_i$ play the role of the average position of the $i$-th particle and are given by \cite{DBBR14,VPVF15,YGP15,YP16}
\begin{align}\label{averagep}
D_i^0=&N!\int dz_1\cdots dz_N\, z_i\theta(z_1<\cdots<z_N)\nonumber\\
&\qquad\qquad\qquad\times\left|\psi_F(\boldsymbol{q}^0)\right|^2\, , \ i=1,\cdots,N\, . 
\end{align}

The factorization of the wavefunctions (\ref{wavef})  simplifies considerably the analytical and numerical investigations of the correlation functions 
in the   strong interacting regime.  A major computational hurdle is represented by the fact that the exchange coefficients (\ref{coeff}), average 
positions  (\ref{averagep}), and, as we will  see in the next sections, the one-body density matrix elements (\ref{obdm}) and the particle densities 
(\ref{singled}) require  the evaluation of $(N-1)$ dimensional integrals. The local exchange coefficients for $N\le 16$  and harmonic trapping  were 
initially  calculated using Monte-Carlo  integration or other approximate methods \cite{DBBR14,VFJV14,VPVF15,LMBP15, YGP15,YP16,YC16,JY16,JY17}.  
Improved numerical algorithms for arbitrary trapping   potentials were introduced in \cite{LKTV16,LKTZ16} for $N\le 35$ and in \cite{DBS16} for $N\le60$ . 
In the case  of harmonic trapping approximate formulae for  the local exchange coefficients using the Local Density Approximation  can be found in  
\cite{YP16,LMBP15} and for the average positions in \cite{YP16}. While  the methods introduced in \cite{LKTV16} and \cite{DBS16} present considerable  
improvements over the previous approaches they require arbitrary precision subroutines  and are time consuming for medium and large numbers of particles. 
In Sec.~\ref{s3} we will develop a more efficient method of computing the relevant charge functions.

\subsection{Excited manifolds and separation of energy scales}\label{s32}

The wavefunctions (\ref{wavef})  and the spin chain (\ref{sc})  describe the groundstate manifold of states of  the strongly interacting spinor gas. In order 
to investigate the temperature correlators or the dynamics of the system  one also  needs to consider the  excited manifolds. The excited manifolds 
are constructed in a similar fashion \cite{YP16} from the excited states 
of the dual system of  spinless fermions,  which will be denoted by $\psi_F(z_1,\cdots,z_N|\boldsymbol{q}^k),\  k=1,2,\cdots$,  and the associated 
spin chain Hamiltonians
\be\label{sce}
H_{sc}^k=E_F(\boldsymbol{q}^k)-\frac{1}{g}\sum_{i=1}^{N-1} J_i^k\left(1+\epsilon \hat{P}_{i,i+1}\right)-G\sum_{i=1}^N D_i^k\sigma_i^z\, ,
\ee
with $E_F(\boldsymbol{q}^k)$ being the energies of the fermionic excited states and $J_i^k$ and $D_i^k$ having similar definitions as in 
(\ref{coeff}) and (\ref{averagep}) but with $\psi_F(\boldsymbol{q}^0)$ replaced by $\psi_F(\boldsymbol{q}^k)$. 
In the case of harmonic trapping the groundstate is $\psi_F=\psi_F(z_1,\cdots,z_N|\boldsymbol{q}_0)$ with $\boldsymbol{q}_0=(0,1,\cdots,N-1)$
while the first excited state is $\psi_F(z_1,\cdots,z_N|\boldsymbol{q}_1)$ with $\boldsymbol{q}_1=(0,1,\cdots,N-2,N)$
and $\phi_j(z)$ the $j$-th Hermite function. Their energies are $E_F(\boldsymbol{q}^0)=\sum_{j=0}^{N-1}(j+1/2)=N^2/2$ and 
$E_F(\boldsymbol{q}^1)=E_F(\boldsymbol{q}^0)+1$, respectively.  For a given sector $\boldsymbol{N}=[N_1,\cdots, N_\kappa]$ and a free 
fermionic state $\boldsymbol{q}$ there are $C^N_{N_1}C^{N-N_1}_{N_2}\cdots C^{N-(N_1+N_2+\cdots+N_{\kappa-2})}_{N_{\kappa-1}}=
N!/[N_1!\cdots N_\kappa!]$ spin states and we will denote  such a state by $|\Phi_{\boldsymbol{N},\boldsymbol{q},n}\rangle$ with 
$n=1,\cdots, N!/[N_1!\cdots N_\kappa!]$.

The explicit form of the effective Hamiltonians (\ref{sc}) and (\ref{sce})  allows for the investigation of the energy scales of the charge and spin sectors. 
In the fermionic case the lower bound of the energy for the spin excitations is obtained by considering the fully antisymmetric  spin  wavefunction $\chi^a$
which satisfies $\hat{P}_{i,i+1}\chi^a=-\chi^a$ obtaining $E_{spin}^{min}=-\frac{2}{g}\sum_{i=1}^{N-1} J_i $. The upper bound  is attained for the fully 
symmetric spin wavefunction $\chi^s$ which satisfies  $\hat{P}_{i,i+1}\chi^s=\chi^s$ resulting in $E_{spin}^{max}=0. $  A similar result can be derived in 
the bosonic case but in this case the upper (lower) bound is attained for the fully antisymmetric (symmetric)  spin wavefunction. If we denote by 
$E(\boldsymbol{N},\boldsymbol{q}^{0,1},n)$  the energies of the $|\Phi_{\boldsymbol{N},\boldsymbol{q}^{0,1},n}\rangle$ eigenstates we have
\begin{align}
E(\boldsymbol{N},\boldsymbol{q}^0,n)=&E_F(\boldsymbol{q}^0)-\frac{1}{g}E_{spin}(\boldsymbol{N},\boldsymbol{q}^0,n)\, ,\ \ \nonumber\\
&|E_{spin}(\boldsymbol{N},\boldsymbol{q}^0,n)|\in\left[0, 2\sum J_i^0\right]\, ,\\
E(\boldsymbol{N},\boldsymbol{q}^1,n)=&E_F(\boldsymbol{q}^1)-\frac{1}{g}E_{spin}(\boldsymbol{N},\boldsymbol{q}^1,n)\, ,\ \  \nonumber\\
&|E_{spin}(\boldsymbol{N},\boldsymbol{q}^1,n)|\in\left[0, 2\sum J_i^1\right]\, ,
\end{align}
where $E_F(\boldsymbol{q^{0,1}})=\sum_{j=1}^N\varepsilon(q_j^{0,1})$ and $E_{spin}(\boldsymbol{N},\boldsymbol{q}^{0,1},n)$ are the energies of the spin states.  
In the strong interaction regime the energy  scale of the spin sector $E_{spin}\sim 1/g$ and the energy scale of the charge sector is $E_{charge}\equiv 
E(\boldsymbol{N},\boldsymbol{q}^1,n)- E(\boldsymbol{N},\boldsymbol{q}^0,n)\sim E_F(\boldsymbol{q}^1)-E_F(\boldsymbol{q}^0)$ with $E_{spin}\ll E_{charge}$. 
These simple considerations show that if we consider temperatures which are much smaller than the energy difference between the groundstate and the first 
excited manifold then we can consider the computation of the correlators over only the groundstate manifold as the contribution from the excited manifolds 
 is negligible.

\subsection{Zero temperature correlators in the strongly interacting regime}

One of the main goals of this paper is to derive  efficient numerical methods for the field-field correlators of strongly interacting  spinor  gases 
at  zero and low temperatures. We are particularly interested in studying the transition from the zero temperature regime which is described by 
LL/bosonization  \cite{H81,G03} in the fermionic case and the ferromagnetic liquid \cite{ZCG07,AT07,MF08,KG09} in the bosonic case to the spin 
incoherent regime \cite{BL,B1,CZ1,CZ2,Matv,FB04,Fiet07}.  We start with the zero temperature correlators. For large but finite repulsion strength 
the Hamiltonian (\ref{sc}) has a unique groundstate (at $g=\infty$ all the spin eigenstates are degenerate) and we will denote the groundstate 
 by $|GS\rangle$. The field-field correlator for the $\sigma$  particles at zero temperature, 
also known as the reduced one-body density matrix,  is defined as
\be\label{corro}
 \rho_\sigma(x,y)=\langle GS|\Psi_\sigma^\dagger(x)\Psi_\sigma(y)|GS\rangle\, ,
\ \sigma=\{1,\cdots,\kappa\}\, .
\ee
In terms of the wavefunction $\psi^{\sigma_1\cdots\sigma_n}(z_1,\cdots,z_N)$  the  field-field correlator can be written as \cite{YGP15}
\begin{widetext}
\be\label{corrwavef}
\rho_\sigma(x,y)=\sum_{\sigma_2,\cdots,\sigma_N=\{1,\cdots,\kappa\}}^{[N_1,\cdots,N_\sigma-1,\cdots,N_\kappa]}
\int dz_2\cdots dz_N\,  \overline{\psi}^{\sigma\sigma_2\cdots\sigma_N}(x,z_2,\cdots,z_N|\boldsymbol{q_0}) 
\psi^{\sigma\sigma_2\cdots\sigma_N}(y,z_2,\cdots,z_N|\boldsymbol{q_0})\, ,
\ee
and in the following it will be sufficient to consider only the case $x\le y$ because from the previous relation we have $\rho_\sigma(y,x)= \overline{
\rho_\sigma(x,y)}$ with the bar denoting complex conjugation. The factorization of the charge and spin degrees of freedom present in the wavefunction 
(\ref{wavef}) can also be made explicit in the expression for the correlator \cite{YGP15,DBS16}
\be\label{sum}
\rho_\sigma(x,y)=\sum_{d_1=1}^N\sum_{d_2=d_1}^N (-\epsilon)^{d_2+d_1} S_\sigma(d_1,d_2)\rho_{d_1,d_2}(x,y)\, ,
\ee
where
\be\label{obdm}
\rho_{d_1,d_2}(x,y)=N!\int_{\Gamma_{d_1,d_2}(x,y)} \prod_{\substack{k=1\\ k\ne d_1}}^N dz_k\, 
\overline{\psi}_F(z_1,\cdots,z_{d_1-1},x,z_{d_1+1}\cdots,z_N|\boldsymbol{q_0}) \psi_F(z_1,\cdots,z_{d_1-1},y,z_{d_1+1}\cdots,z_N|\boldsymbol{q_0})\, ,
\ee
with $\Gamma_{d_1,d_2}(x,y)={L_-\le z_1<\cdots< z_{d_1-1}<x<z_{d_1+1}<\cdots<z_{d_2}<y<z_{d_2+1}<\cdots<z_N\le L_+}$. Here $L_{\pm}$ are the boundaries 
of the  system, in the case of harmonic trapping we have $L_{\pm}=\pm \infty$ while in the case of Dirichlet boundary conditions $L_{\pm}=\pm L/2$.
\end{widetext}
While the one-body density matrix elements $\rho_{d_1,d_2}(x,y)$ are independent on $\sigma$ the spin functions are defined as 
\begin{align}\label{defspin}
S_\sigma (d_1,d_2)
&=\langle\chi|P_\sigma^{(d_1)} (d_1\cdots d_2)|\chi\rangle\, ,
\end{align}
where $P_\sigma^{(d_1)}=|\sigma\rangle_{d_1}\langle \sigma|_{d_1}$ is the projector that selects the states that have a $\sigma$ particle at position 
$d_1$  and $(d_1\cdots d_2)$ is the permutation that cyclically permutes the spins between the positions $d_1$ and $d_2$ 
\be
(d_1\cdots d_2)=\left(\begin{array}{ccccccc}  \cdots & d_1 & d_1+1 & \cdots & d_2-1&  d_2 &\cdots  \\
 \cdots & d_2 & d_1 & \cdots & d_2-2&  d_2-1 &\cdots  
\end{array}\right) .
\ee
From the correlators (\ref{corro}) we can obtain the densities $\rho_\sigma(x)\equiv \rho_\sigma(x,x),$  mean occupation numbers  $\rho_\sigma(n)=\int
\int \phi_n(x)\bar{\phi}_n(y)\rho_\sigma(x,y)\, dxdy\, ,$ and the momentum distributions
\be
n_\sigma(k)=\frac{1}{2\pi}\int\int e^{-i k(x-y)}\rho_\sigma(x,y)\, dxdy .
\ee 
A factorized expression for the densities can be derived from  (\ref{sum}) evaluated at $x=y$ with the result
\be\label{singlesum}
\rho_\sigma(x)=\sum_{d=1}^N S_\sigma(d)\rho_d(x)\, ,
\ee
where
\begin{align}\label{singled}
\rho_{d}(x)&=N!\int_{\Gamma_{d}(x)} \prod_{\substack{k=1 \\ k\ne d}}^N  dz_k\,\nonumber\\ 
&\ \times\left|\psi_F(z_1,\cdots,z_{d-1},x,z_{d+1}\cdots,z_N|\boldsymbol{q}^0)\right|^2,
\end{align}
$\Gamma_{d}=L_-\le z_1<\cdots z_{d}<x<z_{d+1}<\cdots<z_N\le L_+ $ and $S_\sigma(d)=\langle\chi| P_\sigma^{(d)} |\chi\rangle$. The quantities (\ref{singled})
are called single particle densities and are related to the average positions of the particles (\ref{averagep})  via $D_i^k=\int z\rho_{d}(z|
\boldsymbol{q}^k)\, dz$.  Similar to the  case  of the local exchange coefficients (\ref{coeff})  the one-body density matrix elements (\ref{obdm}) 
and  single particle densities (\ref{singled}) require the computation of $(N-1)$-dimensional integrals over products of Slater determinants which are 
not easy to evaluate. In addition to Monte-Carlo integration other methods to compute (\ref{obdm}) and (\ref{singled}) use Chebyshev  interpolation 
\cite{DBS16}  and the connection between  the one-body density matrix elements  and the correlation functions of impenetrable anyons \cite{YP17,JY18}.

\subsection{Finite temperature correlators in the strong interacting regime}

In order to investigate the transition between the LL/ferromagnetic liquid and SILL regime we will consider the correlation functions at temperatures
that are large enough to completely excite the spin sector  but smaller than $E(\boldsymbol{N},\boldsymbol{q}^1,n)- E(\boldsymbol{N},\boldsymbol{q}^0,n)
\sim E_F(\boldsymbol{q}^1)-E_F(\boldsymbol{q}^0)$   such that only the first manifold of states contribute 
to the thermal trace (see the discussion in Sec.~\ref{s32}).
We use units of $k_B=1$ where $k_B$ is the Boltzmann constant. In the canonical ensemble the low temperatures correlators in the $\boldsymbol{N}=[N_1,\cdots,
N_\kappa]$ sector  are [$\sigma=\{1,\cdots,\kappa\}$]
\begin{align}\label{corrt}
\rho_\sigma^T(x,y)=&\sum_{n=1}^{N!/[N_1!\cdots N_\kappa!]} \frac{e^{-E(\boldsymbol{N},\boldsymbol{q}^0,n)/T}}{Z} \nonumber\\
&\ \ \ \times \langle\Phi_{\boldsymbol{N},\boldsymbol{q}^0,n}|\Psi_\sigma^\dagger(x)\Psi_\sigma(y)|\Phi_{\boldsymbol{N},\boldsymbol{q}^0,n}\rangle \, ,
\end{align}
with $Z=\sum_{n=1}^{N!/[N_1!\cdots N_\kappa!]} e^{-E(\boldsymbol{N},\boldsymbol{q}^0,n)/T}$. In the limit $T\rightarrow 0$ the finite temperature correlator 
(\ref{corrt}) reduces to (\ref{corro}). At temperatures for which the thermal energy is much larger than the spin energy of the first manifold $|E_{spin}
(\boldsymbol{N},\boldsymbol{q}^0,n)|/g\ll T$ but smaller than the energies of the first excited manifold $T\ll |E(\boldsymbol{N},\boldsymbol{q}^1,n)|$  we 
have $e^{-E(\boldsymbol{N},\boldsymbol{q}^0,n)/T}\sim e^{-E_F(\boldsymbol{q}^0)/T}$ and (\ref{corrt}) becomes
\begin{align}\label{corrsill1}
\rho_\sigma^{SILL}(x,y)=&\sum_{n=1}^{N!/[N_1!\cdots N_\kappa!]} \frac{1}{Z} \nonumber\\
&\  \times \langle\Phi_{\boldsymbol{N},\boldsymbol{q}^0,n}|\Psi_\sigma^\dagger(x)\Psi_\sigma(y)|\Phi_{\boldsymbol{N},\boldsymbol{q}^0,n}\rangle \, ,
\end{align}
with $Z=N!/[N_1!\cdots N_\kappa!]$ which defines the spin-incoherent Luttinger liquid correlator.  It is important to note that while the spin sector is 
effectively at infinite temperature the charge sector is still in the low temperature regime.  All these considerations were made in the case of large but 
finite repulsion strength. In the limiting case of impenetrable particles $g=\infty$  all the spin states are degenerate and we can have the system in the 
SILL regime even at zero temperature \cite{BL,B1,CZ1,CZ2,Fiet07}.

\section{Evaluation of multidimensional integrals}\label{s3}

In this section we are going to develop a new method of evaluating the multidimensional integrals that appear in the definitions of the 
local exchange coefficients (\ref{coeff}), one-body density matrix elements (\ref{obdm}) and the single particle densities (\ref{singled}).
The main ingredient of our method is the so-called ``phase trick" \cite{IP98,ID06,Patu22} which helps express  multidimensional integrals over 
irregular domains  of $\mathbb{R}^{N-1}$ as Fourier type integrals over determinants of matrices constructed from partial overlaps of the 
single  particle orbitals. The Fourier integrals need to be evaluated only for a discrete set of points task which can be easily accomplished 
using the Discrete Fourier Transform.

\subsection{The single particle densities}\label{spd}

The single particle densities are the simplest of the charge functions involving $N-1$ dimensional integrals. As we will see their treatment
contains all the necessary techniques required to investigate the more complicated functions. We will consider first the case in which the 
eigenstate is in the groundstate manifold. The generalization for the case of excited manifolds will be presented at the end of the section. 

In Appendix \ref{app1} it is shown that the expression for the single particle densities Eq.~(\ref{singled}) can be put in a Fourier integral 
form
\begin{align}\label{a13}
\rho_d(x)&=\int_{0}^{2\pi}\frac{d\alpha}{2\pi}e^{-i(d-1)\alpha}\int_{0}^{2\pi}\frac{d\beta}{2\pi}e^{-i\beta}\nonumber\\
&\qquad\qquad\times\det_{N}\left[e^{i\alpha} \textsf{M}^0+e^{i\beta}\textsf{M}^r+\textsf{M}^1\right](x)\, ,
\end{align}
where $\textsf{M}^{0,1,r}(x)$ are three  $N\times N$ matrices with elements
\begin{subequations}
\begin{align}
[\textsf{M}^0(x)]_{a,b}&=\int_{L_-}^x \overline{\phi}_a(z)\phi_b(z)\, dz\, , \label{defm0}\\
[\textsf{M}^1(x)]_{a,b}&=\int_{x}^{L_+} \overline{\phi}_a(z)\phi_b(z)\, dz\, ,\label{defm1}\\
[\textsf{M}^r(x)]_{a,b}&=\overline{\phi}_a(x)\phi_b(x)\, .\label{defmr}
\end{align}
\end{subequations}
This expression  can be simplified even further. $\textsf{M}^r(x)$ is a rank $1$ matrix and can be written as $\textsf{M}^r(x)=u v^T$ with 
$u=\left(\overline{\phi}_1(x),\cdots,\overline{\phi}_N(x)\right)^T$ and $v=\left(\phi_1(x),\cdots,\phi_N(x)\right)$. Now we can use a 
theorem which states that for an arbitrary matrix $\textsf{M}$ and a matrix of rank 1, which can always can be written as  $u v^T$ with 
$u,v$ arbitrary column vectors, the following identity holds \cite{M90}
\be\label{identity}
\det(\textsf{M}+u v^T)=\det \textsf{M}+ v^T\mbox{adj}(\textsf{M})\, u\, ,
\ee
with $ \mbox{adj}(\textsf{M})=\det(\textsf{M})\textsf{M}^{-1}$ the adjugate matrix of $\textsf{M}$. Using this identity with $\textsf{M}=
e^{i\alpha}  \textsf{M}^0+\textsf{M}^1$ we obtain $\det_{N}\left(e^{i\alpha} \textsf{M}^0+e^{i\beta}\textsf{M}^r+\textsf{M}^1\right)=\det_N
\textsf{M}+e^{i\beta} v^T\mbox{adj} (\textsf{M})\, u$. In (\ref{a13}) the integration over $\beta$ selects the term $v^T\mbox{adj}
(\textsf{M})\, u$ which is in fact $\det_{N}\left(e^{i\alpha} \textsf{M}^0+\textsf{M}^1+\textsf{M}^r\right)-\det_{N}\left(e^{i\alpha} 
\textsf{M}^0+\textsf{M}^1\right)$. Therefore, we find
\begin{align}\label{a14}
\rho_d(x)&=\int_{0}^{2\pi}\frac{d\alpha}{2\pi}e^{-i(d-1)\alpha}
\left[\det_{N}\left(e^{i\alpha} \textsf{M}^0+\textsf{M}^1+\textsf{M}^r\right)\right.\nonumber\\
&\left.\qquad\qquad\qquad -\det_{N}\left(e^{i\alpha} 
\textsf{M}^0+\textsf{M}^1\right)\right](x)\, .
\end{align}
In the literature there are powerful numerical  algorithms \cite{I02,BS91} to deal with this type of Fourier integrals but we can be
more efficient by noticing that we need to compute (\ref{a14}) only for $d=1,\cdots,N$ for a given $x$.  Let 
\begin{align}\label{falpha}
f(\alpha|x)=&\left[\det_{N}\left(e^{i\alpha} \textsf{M}^0+\textsf{M}^1+\textsf{M}^r\right)\right.\nonumber\\
&\qquad\qquad\qquad\left. -\det_{N}\left(e^{i\alpha}  \textsf{M}^0+\textsf{M}^1\right)\right](x)\, ,
\end{align}
which is a polynomial of order $N-1$ in $e^{i\alpha}$ (that it is not of order $N$ can be seen either from (\ref{a11}) or using the 
identity  (\ref{identity}) and the fact that the adjugate is the transpose of its cofactor matrix) and can be written as $f(\alpha|x)=
\sum_{n=0}^{N-1} a_n(x)e^{i\alpha n}$ with $a_n(x)=\rho_{n+1}(x)$. Computing the coefficients of this polynomial (or, equivalently, the 
single particle  densities) can  be done as  follows. First, we evaluate $f_k\equiv f\left(\frac{2\pi k}{N}|x\right)=\sum_{n=0}^{N-1}a_n(x) 
e^{i\frac{2\pi k}{N} n}$ for $k=0,1,\cdots,N-1$. 
Having computed the $N$ vector $\boldsymbol{f}\equiv (f_0,\cdots,f_{N-1})$ then the coefficients are obtained as 
\be
a_n(x)\equiv \rho_{n+1}(x)=\frac{1}{N}\sum_{k=0}^{N-1}f_k(x) e^{-i \frac{2\pi n}{N} k}\, ,
\ee
which is in fact the inverse Discrete Fourier Transform of the vector $\boldsymbol{f}$.

Summarizing, the algorithm for the calculation of the single particle densities (\ref{singled}) is the following: i) compute the matrices 
$\textsf{M}^{0,1,r}(x)$ defined in (\ref{defm0}), (\ref{defm1}) and (\ref{defmr}) (note that in many cases of interest like harmonic trapping, 
Dirichlet or Neumann boundary conditions  the overlaps of single particle orbitals can be analytically computed \cite{P91,AGBK17}) and even 
when this is not the case the numerical evaluation of 1D  integrals is moderately computationally expensive); ii) construct the vector 
$\boldsymbol{f}$  by evaluating (\ref{falpha}) at $2\pi k/N\, , k=0,1,\cdots,N-1$; iii) perform an inverse Discrete Fourier Transform on the 
vector $\boldsymbol{f}$.

In the case of single particle densities for higher excited manifolds which are described by $\psi_F(z_1,\cdots,z_N|\boldsymbol{q})$  all the 
results derived in this section remain valid  but in this case the definitions of the $\textsf{M}^{0,1,r}$ matrices now become 
$[\textsf{M}^0(x)]_{a,b}=\int_{L_-}^x \overline{\phi}_{q_a}(z)\phi_{q_b}(z)\, dz$,  $[\textsf{M}^1(x)]_{a,b}=\int_{x}^{L_+} 
\overline{\phi}_{q_a}(z) \phi_{q_b}(z)\, dz$, and $[\textsf{M}^r(x)]_{a,b}=\overline{\phi}_{q_a} (x)\phi_{q_b}(x)$.

\subsection{Local exchange coefficients}\label{lec}

The computation of the local exchange coefficients (\ref{coeff}) follows along the same lines as in the case of single particle densities
with the additional complication of dealing with another integration. In Appendix \ref{app2} we show that the expression (\ref{coeff}) can 
be written as 
\be\label{c3}
J_d^0=\int_{L_-}^{L_+}  I_d(\xi)\, d\xi\, ,
\ee\, 
with 
\begin{align}\label{c1}
I_d(\xi)&=\int_{0}^{2\pi}\frac{d\alpha}{2\pi}e^{-i(d-1)\alpha} \int_{0}^{2\pi}\frac{d\beta}{2\pi}e^{-i\beta} \int_{0}^{2\pi}
\frac{d\gamma}{2\pi}e^{-i\gamma} \nonumber\\
&\ \ \times \det_{N}\left[e^{i\alpha}\textsf{M}^0+e^{i\beta}\textsf{M}^r+e^{i\gamma}\textsf{M}^d +\textsf{M}^1\right](\xi)\, .
\end{align}
We introduced a new $N\times N$  matrix defined by [$\phi_a'(\xi)$ is the derivative of $\phi_a(\xi)$]
\begin{align}
[\textsf{M}^d(\xi)]_{a,b}&=\overline{\phi'}_a(\xi)\phi'_b(\xi)\, ,\label{defmdj}
\end{align}
and the matrices $\textsf{M}^{0,1,r}(\xi)$ are defined in (\ref{defm0}), (\ref{defm1}), (\ref{defmr}). Both $\textsf{M}^d(\xi)$ and 
$\textsf{M}^r(\xi)$ are rank 1 matrices and the integrations over $\beta$ and $\gamma$ can be eliminated like in  Sec.~\ref{spd} by using 
the  identity (\ref{identity}) twice obtaining 
\begin{widetext}
\begin{align}\label{a17}
I_d(\xi)&=\int_{0}^{2\pi}\frac{d\alpha}{2\pi}e^{-i(d-1)\alpha} 
\left[
\det_N\left(e^{i\alpha}\textsf{M}^0+\textsf{M}^1+\textsf{M}^r+\textsf{M}^d\right)
-\det_N\left(e^{i\alpha}\textsf{M}^0+\textsf{M}^1+\textsf{M}^d\right)\right.\nonumber\\
&\qquad\qquad\qquad\qquad\qquad\qquad\qquad\qquad\qquad\left.
-\det_N\left(e^{i\alpha}\textsf{M}^0+\textsf{M}^1+\textsf{M}^r\right)
+\det_N\left(e^{i\alpha}\textsf{M}^0+\textsf{M}^1\right)
\right](\xi)
\, .
\end{align}

Like in the previous section we do not need to compute the integral to obtain $I_d(\xi)$. Let
\begin{align}\label{c4}
g(\alpha|\xi)&=
\left[\det_N\left(e^{i\alpha}\textsf{M}^0+\textsf{M}^1+\textsf{M}^r+\textsf{M}^d\right)
-\det_N\left(e^{i\alpha}\textsf{M}^0+\textsf{M}^1+\textsf{M}^d\right)\right.\nonumber\\
&\qquad\qquad\qquad\qquad\qquad\qquad\qquad\qquad\ \ \ \left.-\det_N\left(e^{i\alpha}\textsf{M}^0+\textsf{M}^1+\textsf{M}^r\right)
+\det_N\left(e^{i\alpha}\textsf{M}^0+\textsf{M}^1\right)\right](\xi)\, ,
\end{align}
\end{widetext}
which is a polynomial of order $N-2$ in $e^{i\alpha}$ i.e $g(\alpha,\xi)=\sum_{n=0}^{N-2} b_n(\xi)e^{i n\alpha}$ with $b_n(\xi)=I_{n+1}(\xi)$. 
Introducing the vector $\boldsymbol{g}(\xi)=(g_0(\xi),\cdots,g_{N-1}(\xi))$ with elements $g_k(\xi)\equiv g(\frac{2\pi k}{N}|\xi)=
\sum_{n=0}^{N-1}b_n(\xi)  e^{i\frac{2\pi k}{N} n}$  then the $I_n(\xi)$  are obtained as the inverse Discrete Fourier Transform of 
$\boldsymbol{g}(\xi)$ 
\be
b_n(\xi)\equiv I_{n+1}(\xi)=\frac{1}{N}\sum_{k=0}^{N-1}g_k(\xi) e^{-i \frac{2\pi n}{N} k}\, .
\ee

The algorithm for computing the local exchange coefficients is as follows: i) approximate the integral (\ref{c3}) with an $M$-point quadrature  
(Chap. IV of \cite{PFTV92}) 
\be\label{c5}
J_d^0=\int_{L_-}^{L_+}  I_d(\xi)\, d\xi \sim \sum_{j=1}^M I_d(\xi_j) w_j\, ,\
\ee
where $\xi_j$ and $w_j$, $j=1,\cdots,M$ are the abscissas and the weights of the quadrature; ii) for each value of the abscissas $\xi_j$ compute 
the matrices $\textsf{M}^{0,1,r,d}(\xi_j)$ defined in (\ref{defm0}), (\ref{defm1}), (\ref{defmr}) and (\ref{defmdj}); iii) compute the vectors 
$\boldsymbol{g}(\xi_j)=(g_0(\xi_j),\cdots,g_{N-1}(\xi_j))$ for each $\xi_j$; iv)  perform a Discrete Fourier Transform on the vectors 
$\boldsymbol{g}(\xi_j)$ to obtain $I_d(\xi_j)$, $d=1,\cdots,N-1$ for each $\xi_j$; v) do the summation in (\ref{c5}). 

The generalization for the case of local exchange coefficients of the  higher excited manifolds which are described by $\psi_F(z_1,\cdots,z_N
|\boldsymbol{q})$   is obtained by changing the definition of the matrices $\textsf{M}^{0,1,r,d}(\xi)$ to  $[\textsf{M}^0(x)]_{a,b}=
\int_{L_-}^x \overline{\phi}_{q_a}(z)\phi_{q_b}(z)\, dz$,  $[\textsf{M}^1(x)]_{a,b}=\int_{x}^{L_+} \overline{\phi}_{q_a}(z)
\phi_{q_b}(z)\, dz$, $[\textsf{M}^r(x)]_{a,b}=\overline{\phi}_{q_a} (x)\phi_{q_b}(x)$  and  $[\textsf{M}^d(\xi)]_{a,b}=\overline{\phi'}_{q_a}
(\xi)\phi'_{q_b}(\xi)$.

\subsection{The one-body density matrix elements}\label{sobdm}

\begin{table*}\centering
\caption{Evaluation times (in seconds) of the local exchange coefficients $J_i$ [Eq.~(\ref{coeff})], single particle densities  $\rho_d(x)$ 
[Eq.~(\ref{singled})] and one-body density matrix elements $\rho_{d_1,d_2}(x,y)$ [Eq.~(\ref{obdm})]  computed using the approach of 
this paper and the method described in \cite{DBS16}.
}\label{tab1}
\ra{1.3}
\begin{ruledtabular}
\begin{tabular}{@{}llcccccccc@{}}
 & & \multicolumn{2}{c}{  $ J_i$  } & \phantom{abc} & \multicolumn{2}{c}{ $\rho_d(x)$}  &
\phantom{abc} & \multicolumn{2}{c}{ $\rho_{d_1,d_2}(x,y)$}  \\
\cmidrule{3-4} \cmidrule{6-7} \cmidrule{9-10}
$N$  & & Sec.~\ref{lec} & Deuretzbacher \textit{et al.} \cite{DBS16} &  & Sec.~\ref{spd} & Deuretzbacher \textit{et al.} \cite{DBS16}&    & Sec.~\ref{sobdm} & Deuretzbacher \textit{et al.} \cite{DBS16} \\ 

\midrule

5    & &   0.07365   & 1.07895           &  &   0.00035  & 0.05158   &  & 0.00139  & 0.16181 \\
10   & &   0.08448   & 3.01429           &  &   0.00056  & 0.12947   &  & 0.00261  & 2.08619 \\
15   & &   0.09863   & 12.7829           &  &   0.00094  & 0.31709   &  & 0.00608  & 14.4611 \\
20   & &   0.11728   & 364.804           &  &   0.00150  & 4.24389   &  & 0.01523  & 742.804 \\
30   & &   0.28917   & 3327.92           &  &   0.00410  & 35.6766   &  & 0.05876  & --      \\
60   & &   1.55943   & -- \footnote{In the code from the arXiv version of \cite{DBS16} an evaluation time of 853895 seconds for $N=60$ is reported for an unspecified 4 core CPU.}    &  &   0.01854  & --        &  & 0.77707  & --       \\
120  & &   28.8309   & --                &  &   0.12357  & --        &  & 12.3981  & -- \\

\end{tabular}
\end{ruledtabular}
\end{table*}

The computation of the one-body density matrix elements (\ref{obdm}) presents some particularities compared  to the case of the   local 
exchange coefficients or the average positions of particles. We will only need to  consider the case $x\le y$  due to the 
fact that $\rho_\sigma(x,y)=\overline{\rho_\sigma(y,x)}$. In Appendix~\ref{app3}  it is shown that (\ref{obdm}) is equivalent to the 
following  Fourier integral expression  
\begin{widetext}
\begin{align}\label{a22}
\rho_{d_1,d_2}(x,y)=\int_{0}^{2\pi}\frac{d\alpha}{2\pi}e^{-i(d_1-1)\alpha} \int_{0}^{2\pi}\frac{d\gamma}{2\pi}e^{-i\gamma}
\int_{0}^{2\pi}\frac{d\beta}{2\pi}e^{-i(d_2-d_1)\beta}
\det_N\left[e^{i\alpha} \textsf{M}^0 +e^{i\gamma} \textsf{M}^n+e^{i\beta} \textsf{M}^2+\textsf{M}^1\right](x,y)\, ,
\end{align}
\end{widetext}
where in addition to $\textsf{M}^{0,1}$ defined in (\ref{defm0}), (\ref{defm1}) we have introduced two $N\times N$ matrices with elements
\begin{subequations}
\begin{align}
[\textsf{M}^2(x,y)]_{a,b}&=\int_x^y \overline{\phi}_a(z)\phi_b(z)\, dz\, , \label{defm2}\\
[\textsf{M}^n(x,y)]_{a,b}&=\overline{\phi}_a(x)\phi_b(y)\, .\label{defmn}
\end{align}
\end{subequations}
The matrix $\textsf{M}^n$ is of rank 1 so we can use the identity (\ref{identity}) to integrate over $\gamma$  obtaining
\begin{align}\label{a23}
\rho_{d_1,d_2}(x,y)&=\int_{0}^{2\pi}\frac{d\alpha}{2\pi}e^{-i(d_1-1)\alpha}\int_{0}^{2\pi}\frac{d\beta}{2\pi}e^{-i(d_2-d_1)\beta}\nonumber\\
&\ \times \left[\det_N\left(e^{i\alpha} \textsf{M}^0 +e^{i\beta} \textsf{M}^2+\textsf{M}^1+ \textsf{M}^n\right)\right.\nonumber\\
&\ \left.-\det_N\left(e^{i\alpha} \textsf{M}^0 +e^{i\beta} \textsf{M}^2+\textsf{M}^1\right)\right](x,y)\, .
\end{align}
Compared with the previous cases now we have a double integral. Let
\begin{align}\label{defhobdm}
h(\alpha,\beta|x,y)=& \left[\det_N\left(e^{i\alpha} \textsf{M}^0 +e^{i\beta} \textsf{M}^2+\textsf{M}^1+ \textsf{M}^n\right)\right.\nonumber\\
&\ \left.-\det_N\left(e^{i\alpha} \textsf{M}^0 +e^{i\beta} \textsf{M}^2+\textsf{M}^1\right)\right](x,y)\, .
\end{align}
As a multivariate polynomial in $e^{i\alpha}$ and $e^{i\beta}$ the maximum degree term  appearing in the expansion of $h(\alpha,\beta|x,y)$ is  
$e^{i n_1\alpha}e^{in_2\beta}$ with $n_1+n_2=N-1$ (this can be seen from (\ref{a21b})). We have 
\be
h(\alpha,\beta|x,y)=\sum_{n_1=0}^{N-1}\sum_{n_2=0}^{N-1} c_{n_1,n_2}(x,y)e^{ i\alpha n_1}e^{i\beta n_2}\, ,
\ee
with $c_{n_1,n_2}(x,y)=0$ if $n_1+n_2>N-1$. Because $c_{d_1-1,d_2-d_1}(x,y)=\rho_{d_1,d_2}(x,y)$ ($d_1,d_2=1,\cdots,N$) this shows that 
$\rho_{d_1,d_2}(x,y)$ is an upper triangular matrix i.e, $\rho_{d_1,d_2}(x,y)=0$ for $d_1<d_2$. Introducing the $N\times N$ matrix $\boldsymbol{h}(x,y)$
with elements ($k_1, k_2=0,\cdots,N-1$)
\be\label{defmh}
\boldsymbol{h}_{k_1,k_2}(x,y)= h\left(\frac{2\pi k_1}{N},\frac{2\pi k_2}{N}|x,y\right)
\, ,
\ee
then the $c_{n_1,n_2}$ coefficients, or, equivalently, the one-body density matrix elements can be obtained via a 2D Discrete Fourier 
Transform
\begin{align}
c_{n_1,n_2}(x,y)=&\frac{1}{N^2}\sum_{k_1=0}^{N-1}\left[e^{-i \frac{2\pi n_1}{N} k_1} \right.\nonumber\\
&\qquad\times\left.\sum_{k_2=0}^{N-1}e^{-i \frac{2\pi n_2}{N} k_2}
\boldsymbol{h}_{k_1,k_2}(x,y)\right]\, .
\end{align}

Therefore, the algorithm for the calculation of the one-body density matrix elements is the following: i) for given $x$ and $y$ 
satisfying $x\le y$ compute the matrices $\textsf{M}^{0,1,2,n}$ defined in (\ref{defm0}), (\ref{defm1}), (\ref{defm2}) and 
(\ref{defmn}); ii) compute the matrix  $\boldsymbol{h}(x,y)$ defined in (\ref{defmh}); iii) the one-body density matrix elements are
obtained by performing a 2D inverse Discrete Fourier Transform on the matrix  $\boldsymbol{h}(x,y)$. In the case of an excited 
manifold described by $\psi_F(z_1,\cdots,z_N|\boldsymbol{q})$ one needs to modify the definition of the matrices $\textsf{M}^{0,1,2,n}$ 
to  $[\textsf{M}^0(x)]_{a,b}= \int_{L_-}^x \overline{\phi}_{q_a}(z)\phi_{q_b}(z)\, dz$, $[\textsf{M}^2(x,y)]_{a,b}= 
\int_x^y\overline{\phi}_{q_a}(z)\phi_{q_b}(z)\, dz$ and so on.

\subsection{Comparison with other approaches}

\begin{figure*}
\includegraphics[width=1\linewidth]{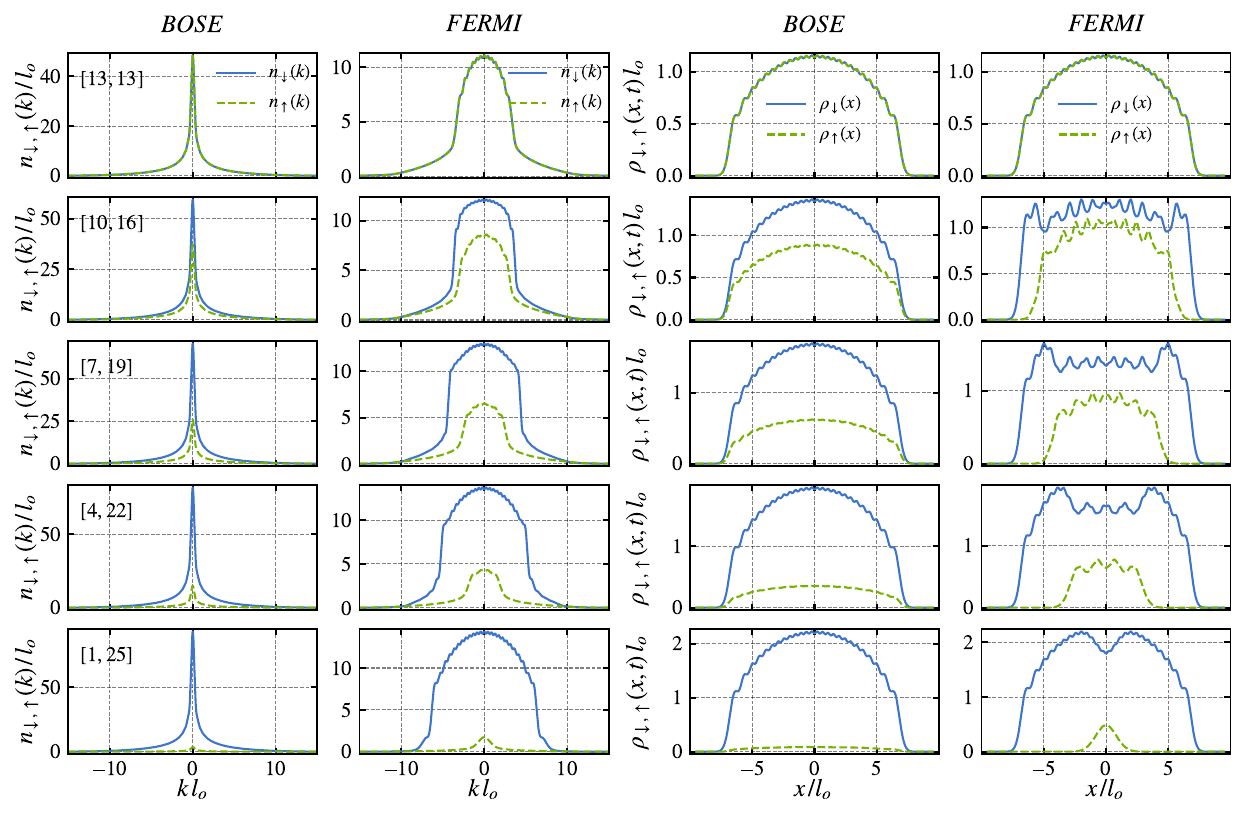}
\caption{Momentum distributions (first and second column) and densities (third and fourth columns) in the groundstates of harmonically 
trapped  two-component bosons and fermions with $N=26$ and different values of population imbalance. First row  $[N_\uparrow,N_\downarrow]
=[13,13]$, second row $[N_\uparrow,N_\downarrow]=[10,16]$, third row $[N_\uparrow,N_\downarrow]=[7,19]$, fourth  row $[N_\uparrow,N_\downarrow]
=[4,22]$ and  fifth row $[N_\uparrow,N_\downarrow]=[1,25]$. The relevant parameters  are $m=\omega=1, g=100$ ($l_o$ is the harmonic oscillator 
length).
}
\label{fig1}
\end{figure*}
\begin{figure}
\includegraphics[width=1\linewidth]{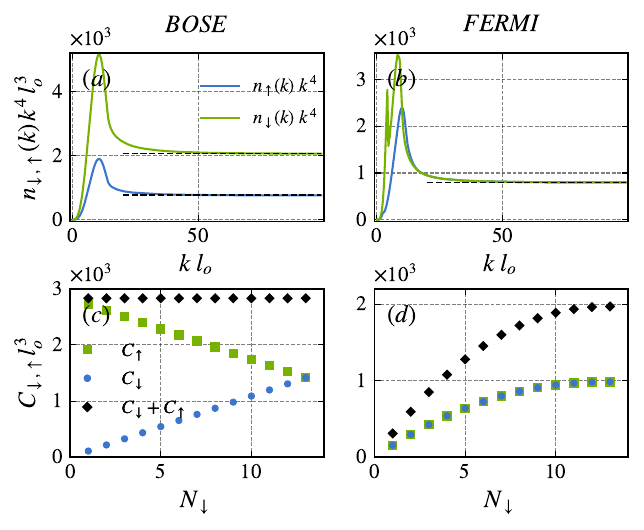}
\caption{Scaled momentum distributions $n_{\uparrow,\downarrow}(k)k^4$ for the groundstate of harmonically trapped bosons a) and fermions b)  with $[N_\uparrow,N_\downarrow]=
[7,19]$ and $m=\omega=1, g=100$.  The dashed  black lines represent the values of the contacts $C_{\uparrow,\downarrow}$. Dependence of 
the contacts on population imbalance for a system  of harmonically trapped bosons c) and fermions d) with  $N=26$. 
}
\label{fig2}
\end{figure}

The algorithms introduced in the previous sections are extremely simple, exact, numerically stable, and do not require the use of arbitrary 
precision  subroutines. The computation of the auxiliary quantities  like the overlap matrices, the determinants and the Discrete Fourier 
Transforms can be done very efficiently using well known techniques \cite{PFTV92}. In Table~\ref{tab1} we present the 
evaluation times of 
the local exchange coefficients, single particle densities and the one-body density matrix elements for harmonically trapped systems with 
different number of particles ranging from $N=5$ to $N=120$ computed using the method introduced in this paper compared with the results 
obtained by running the code provided in the arXiv version of \cite{DBS16}. The results reported were computed using  an AMD 
processor (Ryzen 9 5900HX at 3.30GHz with 8 cores) and  64 GB of RAM using the same interpreted language as in \cite{DBS16}. In order to 
simulate the case of an arbitrary potential we have calculated the overlap matrices using the Clenshaw-Curtis quadrature with  $10$ to $20$ 
points per unit of harmonic oscillator length $l_o=\sqrt{\hbar/(m\omega)}$.
For all values of $N$ and especially for medium and large number of particles our method significantly outperforms the approach of 
\cite{DBS16}. This statement is also true in the case of the method introduced in \cite{LKTV16} in which the authors report that the 
calculation of the local exchange coefficients on an Intel Xenon processor (E5-2630 v3 at 2.40GHz with 8 cores) for $N=10$ takes approximately  
10 seconds, for $N=20$ less than 10 minutes and about an hour for $N=30$. 
The one-body density matrix elements can also be evaluated using the method introduced in \cite{YP17}. In the general 
case  it requires the computation of $\sim N^4$ overlap integrals of anyonic type and $\sim N^4$ determinants of $N\times N$ matrices. Our 
method  requires only $\sim N^2$ integrals and $\sim N^2$ determinants which represents a large polynomial improvement in numerical efficiency.

\vspace{1.5cm}

\section{Numerical results}\label{s4}

The method introduced in the previous section  substantially reduce the computation time of the correlators  allowing for the investigation of 
systems with a larger number of particles than before. In  Fig.~\ref{fig1} we present results for the densities and momentum 
distributions of a harmonically trapped two-component  system of $N=26$  particles with $g=100$ and different values of the population 
imbalance. We mention that the computational limitation in this case comes from the exact diagonalization of the effective spin chain (\ref{sc}) 
with $D_i^0=0$ which in the balanced sector $N_\downarrow=N_\uparrow=13$ has dimension $C^{26}_{13}=10400600$. Employing another method, like 
DMRG, for the calculation of the spin functions $S_\sigma(d_1,d_2)$  $(\ref{defspin})$  one could in principle study systems with up to 100
particles. In the balanced case the densities are the same for both statistics and are very close to the one for noninteracting  spinless fermions 
$\rho_\sigma(x)\sim\frac{1}{2}\rho_{FF}(x)$  with $\rho_{FF}(x)=\frac{1}{2}\sum_{j=0}^{N}|\phi_{j}(x)|^2$. As  functions of the population 
imbalance the bosonic densities satisfy $\rho_{\sigma}(x)\sim (N_\sigma/N)\rho_{FF}(x)$ (third column of Fig.~\ref{fig1}) while in the fermionic 
case the density  profiles reorganize such that the spin-up and spin-down parts avoid overlapping \cite{GCWM09}. While the overall 
shape of the total density profiles (ignoring the small oscillations) can be obtained using the Thomas-Fermi approximation, this approach does not 
produce the correct density profiles of individual components in the fermionic case.

The momentum distributions for bosons have the characteristic shape of a quasicondensate with a large number of particles with  momenta close to 
$k=0$ while  in the case of fermions they are  similar to the momentum  distribution of spinless noninteracting 
fermions above a flat background \cite{DBS16}.  
Some of the previous numerical investigations of the momentum distribution for trapped and homogeneous multicomponent systems can be found in \cite{OS90,DVLR15,YGP15,DBS16,DJAR16,JY16,JY17,JY18,PV20,OPCA23,COAP23}.
A distinctive feature of the fermionic distribution is the presence of small oscillations with the 
number of maxima equaling the number of particles  in each component.  In both cases the momentum distributions have wide tails and for large values 
of $k$ they 
behave like  $\lim_{k\rightarrow \infty} n_\sigma(k)\sim C_\sigma/k^4$  with $C_\sigma$ the Tan contacts \cite{Tan08a,Tan08b,Tan08c,OD03,VZM12,
BZ11,PK17}. The $1/k^4$ tail is an universal feature of systems with contact interactions and the contacts are experimentally measurable. While the 
total contact $C=\sum_{\sigma=1}^\kappa  C_\sigma$ can be obtained from the groundstate energy of the spin chain (\ref{sc})  the contacts for each 
component  need to be computed using other methods. In Fig.~\ref{fig2}c) and \ref{fig2}d) we present the dependence on population 
imbalance of  the contacts for each component, $C_\sigma$, and total contact, $C$, for a trapped two-component system with  $N=26$. The contacts are 
obtained by fitting the tails of the momentum distributions [see Fig.~\ref{fig2}a) and \ref{fig2}b)].
In the bosonic case  $C$ is independent of imbalance but the individual contacts increase and decrease linearly as a function of 
$N_\downarrow$. This is to be expected when we take into account that we consider the inter- and intra-particle interactions to be equal and that 
the contacts are directly proportional with the interaction energy. In the fermionic case the individual contacts are equal, for an analytical proof 
see \cite{BZ11,PK17}, and they are a monotonically increasing, but not linearly, function of $N_\downarrow$ with the maximum 
attained for the balanced system. For the same value of imbalance the total contact of the bosonic system is larger than the fermionic one 
resulting in a wider momentum distribution in the tails.

\begin{figure*}
\includegraphics[width=1\linewidth]{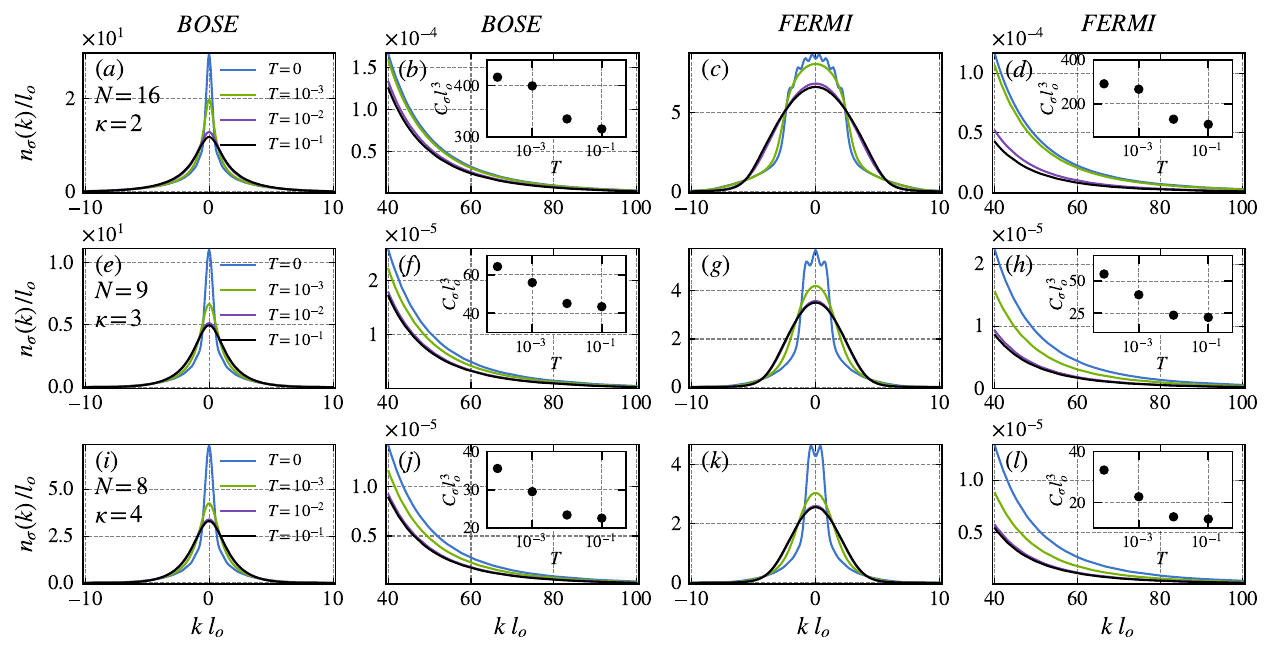}
\caption{Dependence on temperature of the momentum distribution for balanced harmonically trapped systems ($m=\omega=1, g=10000$) with $N=16, \kappa=2$ 
(first row),  $N=9, \kappa=3$ (second row) and $N=8,\kappa=4$ (third row). The second and fourth column contains zooms of the right tails of the 
momentum distributions in the interval $k\in[40,100]$ with the insets depicting the Tan contacts computed at $T=\{10^{-4},10^{-3},10^{-2},10^{-1}\}\times 
 \omega$. The values of the contacts at  $T=10^{-4}\omega$ are almost indistinguishable from the zero temperature contacts.
}
\label{fig3}
\end{figure*}

The role of the temperature in strongly interacting spinor gases has not been sufficiently explored in the literature. This is a rather remarkable 
oversight  when we take into account that  strongly interacting systems with internal degrees of freedom present two temperature scales and small 
changes in the  temperature can be accompanied by dramatic changes in their static and dynamic properties \cite{CSZ05,Fiet07}.  
We investigate first  the dependence on temperature  of the momentum distribution using Eq.~(\ref{corrt}) for the correlators. We  consider 
temperatures  ranging from zero to $E_{spin}\ll T \ll E_{charge}$ that cover  the transition from the LL/ferromagnetic liquid phase to the SILL 
regime.
In Fig.~\ref{fig3} we present the temperature dependence of the momentum distribution for balanced  harmonically trapped systems with strong 
interactions, $g=10000$, and  $\kappa=\{2,3,4\}$. For the largest temperature considered $T=0.1\omega$  all the systems are in the SILL regime 
as the  spin sector is almost completely excited. We see that as the temperature is increased, contrary to the usual expectations, the number of 
particles at high momenta decreases, feature which is best  exemplified by the monotonically decreasing contacts (see the insets in the second and 
fourth column of Fig.~\ref{fig3}). This momentum reconstruction is accompanied by a decrease of the number of particles at momenta close to zero and 
an increase in the number of particles at intermediate momenta. This remarkable phenomenon is within reach of current ultracold gases experiments 
\cite{PMCL14,CSKH23} and was predicted for two-component systems \cite{CSZ05,PKF18,PV20}. Here, in addition to mapping the entire transition, we show that 
it is present in  systems with more than two components being a general feature of multicomponent systems. In \cite{PV20} it was argued that the 
minimum of the contact is due to the mixing of states with different exchange symmetries and we expect that the amplitude of the reconstruction to 
decrease as the number of components becomes large. At higher temperatures, when the charge sector also becomes excited, the contacts 
and the tails behave conventionally increasing with temperature \cite{DJAR16}.

\begin{figure}
\includegraphics[width=0.97\linewidth]{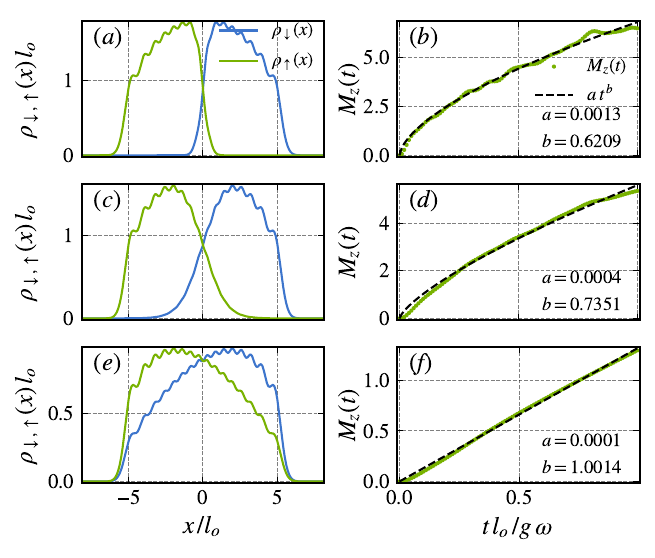}
\caption{Time and temperature dependence (first row $T=0$, second row $T=0.01\omega$  and third row $T=0.1\omega$) of the integrated magnetization $M_z(t)$ 
for a balanced two-component fermionic system with $N=16$ particles after a quench from a  domain wall like state. The parameters of the initial Hamiltonian (\ref{sc}) 
are $G=0.014, g=10000$ and $m=\omega=1$. In the first column we present the initial densities at $t=0$ and in the second column $M_z(t)$ (continuous green line) and 
the best fit obtained for $a\,t^b$ (dashed black line) with $a$ and $b$ free parameters. 
}
\label{fig4}
\end{figure}

Small changes in temperature can also produce impressive changes  in the transport properties of strongly interacting spinor gases. In Fig.~\ref{fig4} we 
present the time dependence of the integrated magnetization 
\be
M_z(t)=\int_{0}^{\infty} \left[\rho_\downarrow(z,t)-\rho_\uparrow(z,t)\right]-\left[\rho_\downarrow(z,0)-\rho_\uparrow(z,0)\right]\, dz\, ,
\ee 
after a quench from a domain wall like state for several values of the temperature. We consider a balanced fermionic system with $N=16$ particles
initially prepared  in an thermal state of (\ref{sc})  with a strong gradient $G$  which results in  spin segregation as it can be seen in the first column 
of Fig.~\ref{fig4}. At $t=0$ we quench the gradient to zero and let the  system evolve. At zero temperature this nonequilibrium scenario  was previously 
investigated in \cite{YGP15,YP16,PVM22}. After the quench the momentum distribution remains almost unchanged (oscillations would be produced if one would 
consider a spinor gas with different inter- and intra-particles couplings like \cite{APPH22} which would break the $SU(2)$ symmetry, see also the discussion 
in \cite{MAMV24}) but the integrated magnetization presents  oscillatory dynamics at large times \cite{PVM22}. At zero temperature and immediately after the 
quench the integrated magnetization presents superdifussive  behaviour  \cite{BGI21,PVM22} with   $M_z(t)\sim t^{0.62}$.  The extremely sensitive nature on 
temperature of the transport properties for multicomponent  systems can be seen in the second column of Fig.~\ref{fig4}. Even minute changes in temperature, 
for which the charge degrees of freedom remain effectively  frozen, produce ballistic transport $M_z(t)\sim t^{1.001}$  as it can be seen in Fig.~\ref{fig4}f).
This result is particularly intriguing because in the case of the homogeneous Hubbard model, which can be understood as the lattice counterpart of the fermionic
two-component spinor gas, it was observed \cite{MWVP23} that after a quench from a thermal state at \textit{infinite temperature} with a weak imbalance in the 
magnetization the spin current presents Kardar-Parisi-Zhang scaling \cite{KPZ86} implying $M_z(t)\sim t^{2/3}$.

\vspace{0.5cm}

\section{Determinant representations for the trapped SILL correlators}\label{s5}

In the spin incoherent regime the factorized nature of the mean value of bilocal operators $\langle\Phi_{\boldsymbol{N},\boldsymbol{q},n}|
\Psi_\sigma^\dagger(x)\Psi_\sigma(y)|\Phi_{\boldsymbol{N},\boldsymbol{q},n}\rangle$ and the method employed in  Sec.~\ref{sobdm} to compute the 
one-body density matrix elements can be used  to derive determinant representations for the SILL correlators at zero  and finite temperature. 
In  this section we will consider the case of impenetrable particles, $g=\infty$, but our results are also true in the case of large and finite 
coupling strength if the thermal energy is much larger than the energy of the spin sector $E_{spin}/g\ll T$.  
When doing numerics at finite temperature, like in Sec.~\ref{s4}, it was preferable to use the canonical ensemble (\ref{corrt}) but in this 
section, focused on analytical derivations, it will be more useful to use the grandcanonical ensemble. In the grandcanonical ensemble at  
temperature $T$  and chemical potentials $\mu_1, \cdots, \mu_\kappa$  an explicit expression for the  field-field correlator in the SILL regime 
is given by (here we consider the sum over all the excited manifolds)
\begin{widetext}
\begin{align}\label{b1}
\rho_\sigma(x,y)&=\frac{1}{Z}\sum_{N=0}^\infty \sum_{q_1<\cdots< q_N} \sum_{N_1=0}^{N}\sum_{N_2=0}^{N-N_1}\cdots 
\sum_{N_{\kappa-1}=0}^{N-(N_1+\cdots+N_{\kappa-2})}
\sum_{n=1}^{N!/[N_1!\cdots N_\kappa!]}\nonumber\\
&\qquad\qquad\qquad\qquad\times e^{-\sum_{j=1}^N\varepsilon(q_j)/T+\sum_{\sigma'=1}^\kappa\mu_{\sigma'} N_{\sigma'}/T}
\langle\Phi_{\boldsymbol{N},\boldsymbol{q},n}|\Psi^\dagger_\sigma(x)\Psi_\sigma(y)|\Phi_{\boldsymbol{N},\boldsymbol{q},n}\rangle
\end{align}
with  $Z$  the grandcanonical partition function (see Appendix~\ref{app4}), $\boldsymbol{N}=[N_1,\cdots,N_\kappa]$ with 
$\sum_{j=1}^\kappa N_j=N$ and  $\boldsymbol{q}=(q_1,\cdots,q_N)$ . For a given $\boldsymbol{N}$ and $\boldsymbol{q}$ there are  $N!/[N_1!\cdots 
N_\kappa!]$ spin eigenstates  which are indexed  by  the subscript $n$ in  $|\Phi_{\boldsymbol{N},\boldsymbol{q},n}\rangle$.  The 
statistical contribution of a spin state is $e^{-E_{spin}(\boldsymbol{N},\boldsymbol{q},n)/gT}$ and because we consider $g=\infty$ or 
temperatures  $E_{spin}/g\ll T$ these factors are effectively equal to $1$. Each mean value of the 
bilocal operators $\langle\Phi_{\boldsymbol{N},\boldsymbol{q},n}|\Psi^\dagger_\sigma(x) \Psi_\sigma(y)|\Phi_{\boldsymbol{N},\boldsymbol{q},n}
\rangle$  appearing in (\ref{b1}) can be written in a charge and spin factorized form like (\ref{sum}) with the result 
\begin{align}\label{b2}
\rho_\sigma(x,y)&=\frac{1}{Z} \sum_{N=0}^\infty \sum_{q_1<\cdots< q_N}  \sum_{d_1=1}^N\sum_{d_2=d_1}^N (-\epsilon)^{d_1+d_2}  
e^{-\sum_{j=1}^N\varepsilon(q_j)/T}
\rho_{d_1,d_2}(x,y|\boldsymbol{q})\, S_{\sigma}^{SILL}(d_1,d_2)\, ,
\end{align}
where 
\begin{align}\label{b3}
 S_{\sigma}^{SILL}(d_1,d_2)&=\sum_{N_1=0}^{N}\sum_{N_2=0}^{N-N_1}\cdots \sum_{N_{\kappa-1}=0}^{N-(N_1+\cdots+N_{\kappa-2})}
\sum_{n=1}^{N!/[N_1!\cdots N_\kappa!]}e^{\sum_{\sigma'=1}^\kappa\mu_{\sigma'} N_{\sigma'}/T}
\langle\chi_{\boldsymbol{N},n}|P_\sigma^{(d_1)} (d_1\cdots d_2)|\chi_{\boldsymbol{N},n}\rangle\, .
\end{align}
\end{widetext}
In (\ref{b2}) $\rho_{d_1,d_2}(x,y|\boldsymbol{q})$ is the generalization of Eq.~(\ref{obdm}) in the case of an excited manifold described by 
$\boldsymbol{q}$. Both $\rho_{d_1,d_2}(x,y|\boldsymbol{q})$ and $S_{\sigma}^{SILL}(d_1,d_2)$ depend on $N$ but we will not make explicit this 
dependence in order to keep the notation light. Even though (\ref{b2}) and (\ref{b3}) seem daunting we will show below that they reduce to very 
simple determinant representations which are easily implementable numerically and can also be used to derive rigorous analytical results.

\subsection{Computation of $ S_\sigma^{SILL}$}

The spin function $S_{\sigma}^{SILL}(d_1,d_2)$ defined above can be understood as the trace of an operator in the Hilbert space of a spin chain 
with $N$  lattice sites and spin $s=(\kappa-1)/2$ 
\begin{align}\label{b4}
 S_{\sigma}^{SILL}(d_1,d_2)&=\mbox{Tr}\left[e^{\sum_{\sigma'=1}^\kappa\mu_{\sigma'} \hat{N}_{\sigma'}/T} P_\sigma^{(d_1)}(d_1\cdots d_2)\right]\, .
\end{align}
Because the trace is invariant to a change of basis it is preferable to use the canonical basis $|\sigma_1\sigma_2\cdots\sigma_N\rangle$ with 
$\sigma_i\in\{1,\cdots,\kappa\}$. For a particular $\kappa$  the cardinality of the basis is $\kappa^N$. The effect of the operator 
$e^{\mu_{\sigma'} \hat{N}_{\sigma'}/T}$ on any element of the basis is simple producing a multiplicative factor $e^{\mu_{\sigma'}/T}$ for any 
element in $|\sigma_1\sigma_2\cdots\sigma_N\rangle$  which is $\sigma'$. The  operator  $P_\sigma^{(d_1)}(d_1\cdots d_2)$ acting on the basis  
selects the vectors that have a particle of type $\sigma$ on position $d_1$ after a cyclic permutation of the spins between the positions $d_1$ 
and $d_2$. Therefore, the only vectors of the basis that give a nonzero contribution to the trace are those that have  spins of type $\sigma$ 
between the $d_1$ and $d_2$ positions \cite{JY18}. Let us consider  the three-component case $\kappa=3$  and $\sigma=1$. The basis elements that 
have only  $l=d_2-d_1+1$ spins of type $\sigma$, all of them between the positions $d_1$  and $d_2$, produce a term $C^{N-l}_0e^{\mu_1 l/T}\left
( \sum_{k=0}^{N-l}C^{N-l}_k e^{\mu_2 k/T}  e^{\mu_3(N-k-l)/T}\right)$ while the basis elements that have $l+1$ spins of type $\sigma$ ($l$ of 
them between $d_1$ and $d_2$) give a contribution  $C^{N-l}_1e^{\mu_1 (l+1)/T}\left( \sum_{k=0}^{N-l-1}C^{N-l-1}_k e^{\mu_2k/T}  
e^{\mu_3(N-k-l-1)/T}\right)$  and so on. Summing all these contributions and using the binomial theorem we obtain $S_1^{SILL}(d_1,d_2)=e^{\mu_1 
l/T}\left(e^{\mu_1 /T} +e^{\mu_2 /T}+e^{\mu_3 /T}\right)^{N-l}$. The obvious generalization for arbitrary $\kappa$ and $\sigma$ is
\begin{align}\label{ssill}
S_{\sigma}^{SILL}(d_1,d_2)=&\left(\frac{e^{\mu_\sigma/T}}{e^{\mu_1/T}+\cdots+ e^{\mu_\kappa /T}}\right)^{d_2-d_1+1}\nonumber\\
&\qquad\times\left(e^{\mu_1/T}+\cdots+ 
e^{\mu_\kappa /T}\right)^N\, .
\end{align}

\subsection{Large $N$ limit of the correlators}

The correlator (\ref{b2}) can also be written as
\begin{align}\label{b2b}
\rho_\sigma(x,y)&=\frac{1}{Z} \sum_{N=0}^\infty \sum_{q_1<\cdots< q_N}  \sum_{d_1=1}^N\sum_{d_2=d_1}^N A_\sigma(d_1,d_2|\boldsymbol{q})\, ,
\end{align}
with 
\begin{align}\label{b5}
A_{\sigma}(d_1,d_2|\boldsymbol{q})=&(-\epsilon)^{d_2-d_1}e^{-\sum_{j=1}^N\varepsilon(q_j)/T} \nonumber\\
&\qquad\times \rho_{d_1,d_2}(x,y|\boldsymbol{q})\, 
S_{\sigma}^{SILL}(d_1,d_2)\, .
\end{align}
The next step in our analysis will take advantage of the extremely simple form of $S_{\sigma}^{SILL}(d_1,d_2)$ (\ref{ssill}) which can be used
to derive a formula similar to (\ref{a23}) for $A_{\sigma}(d_1,d_2|\boldsymbol{q})$ . We can do this because by 
multiplying a matrix appearing in (\ref{a23}) with a  constant $c$ then,  in the final result, we will have: a factor of  $c^{d_1-1}$ if we 
multiply  $\textsf{M}^0$, a  factor of $c$ if we multiply $\textsf{M}^n$, a  factor of $c^{d_2-d_1}$ if we multiply $\textsf{M}^2$, and a factor 
of $c^{N-d_2}$ if we multiply  $\textsf{M}^1$. We introduce [$a=1,\cdots,N$]
\begin{align}
\tilde{\vartheta}_a(\boldsymbol{q})&=e^{-\varepsilon(q_a)/T}\left(e^{\mu_1/T}+\cdots+e^{\mu_\kappa/T}\right)\, ,\label{defhfunc}\\
f(\sigma)&=\left(\frac{e^{\mu_\sigma/T}}{e^{\mu_1/T}+\cdots+ e^{\mu_\kappa /T}}\right)\, , \ \ \sigma=\{1,\cdots,\kappa\}\label{deffunc}
\end{align}
and four $N\times N$ matrices dependent on $\boldsymbol{q}$  with elements 
\begin{subequations}
\begin{align}
&[\tilde{\textsf{M}}^0(x)]_{a,b}=\tilde{\vartheta}_a^{1/2}(\boldsymbol{q})\left(\int_{L_-}^x \overline{\phi}_{q_a}(z)\phi_{q_b}(z)\, dz\right)
\tilde{\vartheta}_b^{1/2}(\boldsymbol{q})\, , \label{deftm0}\\
&[\tilde{\textsf{M}}^1(y)]_{a,b}=\tilde{\vartheta}_a^{1/2}(\boldsymbol{q})\left(\int_{y}^{L_+}\overline{\phi}_{q_a}(z)\phi_{q_b}(z)\, dz\right)
\tilde{\vartheta}_b^{1/2}(\boldsymbol{q})\, , \label{deftm1}\\
&[\tilde{\textsf{M}}^2(x,y)]_{a,b}=\tilde{\vartheta}_a^{1/2}(\boldsymbol{q})\left(\int_{x}^{y}\overline{\phi}_{q_a}(z)\phi_{q_b}(z)\, dz\right)
\tilde{\vartheta}_b^{1/2}(\boldsymbol{q})\, , \label{deftm2}\\
&[\tilde{\textsf{M}}^n(x,y)]_{a,b}=\tilde{\vartheta}_a^{1/2}(\boldsymbol{q})\left(\overline{\phi}_{q_a}(x)\phi_{q_b}(y)\right)
\tilde{\vartheta}_b^{1/2}(\boldsymbol{q})\, . \label{deftmn}
\end{align}
\end{subequations}
Then 
\begin{align}
A_{\sigma}(d_1,d_2|\boldsymbol{q})=&\int_{0}^{2\pi}\frac{d\alpha}{2\pi}e^{-i(d_1-1)\alpha}\nonumber\\
&\ \ \times\int_{0}^{2\pi}\frac{d\beta}{2\pi}e^{-i(d_2-d_1)\beta} 
p_{\sigma}(\alpha,\beta|\boldsymbol{q})\, ,
\end{align}
with 
\begin{align}\label{b6}
p_{\sigma}(\alpha,\beta|\boldsymbol{q})=&\det_N\left(e^{i\alpha}\tilde{\textsf{M}}^0 -\epsilon f(\sigma) e^{i\beta} \tilde{\textsf{M}}^2+ 
\tilde{\textsf{M}}^1+ f(\sigma) \tilde{\textsf{M}}^n\right)\nonumber\\
&\ -\det_N\left(e^{i\alpha}\tilde{\textsf{M}}^0 -\epsilon f(\sigma) e^{i\beta} \tilde{\textsf{M}}^2+ \tilde{\textsf{M}}^1\right)\, .
\end{align}
Similar to the case of the function $h(\alpha,\beta)$ defined in (\ref{defhobdm}) and analyzed in Sec.~\ref{sobdm} $p_{\sigma}(\alpha,\beta|
\boldsymbol{q})$is a multivariate polynomial in $e^{i\alpha}$ and $e^{i\beta}$ with the maximum degree term given by  $e^{i n_1\alpha}
e^{in_2\beta}$ with $n_1+n_2=N-1$.  We have 
\begin{align}
p_\sigma(\alpha,\beta|\boldsymbol{q})=&\sum_{n_1=0}^{N-1}\sum_{n_2=0}^{N-1} a_{n_1,n_2}(\boldsymbol{q})e^{ i\alpha n_1}e^{i\beta n_2}\, ,
\end{align}
with $a_{n_1,n_2}(\boldsymbol{q})=0$ if $n_1+n_2>N-1$. Because $a_{d_1-1,d_2-d_1}(\boldsymbol{q})=A_{\sigma}(d_1,d_2|\boldsymbol{q})$ ($d_1,d_2=1,
\cdots,N$)the following identity holds
\be
\sum_{d_1=1}^N\sum_{d_2=d_1}^N A_\sigma(d_1,d_2|\boldsymbol{q})=\sum_{n_1=0}^{N-1}\sum_{n_2=0}^{N-n_1-1}a_{n_1,n_2}(\boldsymbol{q})\equiv 
p_{\sigma}(0,0|\boldsymbol{q})\, ,
\ee
and from (\ref{b2}) we obtain
\be
\rho_\sigma(x,y)=\frac{1}{Z} \sum_{N=0}^\infty\sum_{q_1<\cdots<q_N} p_{\sigma}(0,0|\boldsymbol{q})\, .
\ee
The summation over $N$ and $\boldsymbol{q}$ can be done with the help of von Koch's determinant formula which states  that for a square matrix $A$ 
of dimension $M$ (which can also be infinite) and $z$ a complex number the following expansion holds
\be\label{koch}
\det(\boldsymbol{1}+z A)=1+z\sum_{m=1}^M A_{m,m}+z^2\sum_{m<n}^M\left|\begin{array}{cc}
A_{m,m} &A_{m,n}\\
A_{n,m} & A_{n,n}
\end{array}
\right|+\cdots\, .
\ee
We find
\begin{align}\label{b7}
\rho_\sigma(x,y)=&\frac{1}{Z} \left[\det\left(\boldsymbol{1}+\tilde{\textsf{M}}^0 -\epsilon f(\sigma)  \tilde{\textsf{M}}^2+ 
\tilde{\textsf{M}}^1+ f(\sigma) \tilde{\textsf{M}}^n\right)\right.\nonumber\\
&\ \  \ \ \left.-\det\left(\boldsymbol{1}+\tilde{\textsf{M}}^0 -\epsilon f(\sigma)  \tilde{\textsf{M}}^2+ \tilde{\textsf{M}}^1\right)\right]\, ,
\end{align}
where now the determinants are infinite and the matrices $\tilde{\textsf{M}}^{0,1,2,n}$ defined in (\ref{deftm0}),  (\ref{deftm1}),  (\ref{deftm2}), 
and (\ref{deftmn}) correspond to the infinite state $\boldsymbol{q}=1,2,\cdots $. The sum of the matrices $\tilde{\textsf{M}}^0$,  
$\tilde{\textsf{M}}^1$ and  $\tilde{\textsf{M}}^2$ can be simplified by noticing that  
\begin{widetext}
\begin{align}
\left[\tilde{\textsf{M}}^0 -\epsilon f(\sigma)  \tilde{\textsf{M}}^2+ \tilde{\textsf{M}}^1\right]_{a,b}&=
\tilde{\vartheta}_a^{1/2}\tilde{\vartheta}_b^{1/2}\left[\left(\int_{L_-}^x -\epsilon f(\sigma) \int_{x}^{y} +\int_{y}^{L_+} \right) 
\overline{\phi}_a(z)\phi_b(z)\, dz \right]\, ,\nonumber\\
&=\tilde{\vartheta}_a^{1/2}\tilde{\vartheta}_b^{1/2}\left\{\left[\left(\int_{L_-}^x + \int_{x}^{y} +\int_{y}^{L_+} \right)- 
(1+\epsilon f(\sigma)) \int_x^y\right] \overline{\phi}_a(z)\phi_b(z)\, dz \right\}\, ,\nonumber\\
&= \tilde{\vartheta}_a^{1/2}\tilde{\vartheta}_b^{1/2}\left[\delta_{a,b} -(1+\epsilon f(\sigma))\int_x^y \overline{\phi}_a(z)\phi_b(z)\, 
dz \right]\, .\label{b8}
\end{align}
\end{widetext}
All that remains to be done is to divide (\ref{b7}) by the partition function $Z=\prod_{q=1}^\infty\left[1+\left(\sum_{\sigma=1}^\kappa  
e^{\frac{\mu_\sigma}{T}}\right) e^{-\varepsilon(q)/T}\right]$ which has been calculated in Appendix~\ref{app4}. Using the result (\ref{b8}) and dividing 
the $a$-th row and column  of the matrices appearing in (\ref{b7}) with   $\left[1+\left(\sum_{\sigma=1}^\kappa e^{\mu_\sigma/T}\right)
e^{-\varepsilon(a)/T}\right]^{1/2}$ we obtain the final result. The  correlators in the SILL regime for an impenetrable spinor gas with $\kappa$ 
components subjected to a trapping potential have the  following representation 
\begin{align}\label{corrsillt}
\rho_\sigma(x,y)=&\det\left[\boldsymbol{1}-(1+\epsilon f(\sigma))\, \textsf{V}_T+f(\sigma) \textsf{R}_T\right]\nonumber\\
&\qquad\qquad\ \ -\det\left[\boldsymbol{1}-(1+\epsilon f(\sigma))\, \textsf{V}_T\right]\, ,
\end{align}
where $\textsf{V}_T$ and $\textsf{R}_T$ are infinite matrices with elements ($a,b=1,2,\cdots$)
\begin{align}
[\textsf{V}_T(x,y)]_{a,b}&=\left(\vartheta(a)\vartheta(b)\right)^{1/2}\int_x^y \overline{\phi}_a(z)\phi_b(z)\, dz\, , \label{defvt}\\
[\textsf{R}_T(x,y)]_{a,b}&=\left(\vartheta(a)\vartheta(b)\right)^{1/2}\overline{\phi}_a(x)\phi_b(y)\, ,\label{defrt}
\end{align}
$f(\sigma)$ is defined in (\ref{deffunc}) and  $\vartheta(a)$ is a generalized Fermi function
\be\label{deffermi}
\vartheta(a)=\frac{\left(e^{\mu_1/T}+\cdots+e^{\mu_\kappa/T}\right)e^{-\varepsilon(a)/T}}{1+\left(e^{\mu_1/T}+\cdots+e^{\mu_\kappa/T}\right)
e^{-\varepsilon(a)/T}}\, .
\ee
The determinant representation (\ref{corrsillt}) is the generalization for arbitrary $\kappa$ of the result derived in \cite{Patu23b} for two-component 
systems. 

We need to make three observations. The first observation is that (\ref{corrsillt}) remains valid in nonequilibrium situations by doing a very simple  
modification.  In the case of impenetrable particles the dynamics is restricted to the charge sector with  the spin sector being frozen \cite{ASYP21,
Patu23a,Patu23b}. This means that out of equilibrium   the spin part of the wavefunction (\ref{wavef}) remains unchanged while the charge  part is 
replaced by (we consider the case of a general manifold characterized by $\boldsymbol{q}$)
\be\label{slatert}
\psi_F(z_1,\cdots,z_N|\boldsymbol{q}; t)=\frac{1}{\sqrt{N!}}\det_N\left[\phi_{q_j}(z_i,t)\right]_{i,j=1,\cdots,N}\, ,
\ee
where $\phi_q(z,t)$ are the evolved single particle orbitals of a spinless fermionic system subjected to the same quench.  Therefore,  the determinant 
representation (\ref{corrsillt}) also describe  nonequilibrium situations if the relevant matrices are being replaced by  $[\textsf{V}_T(x,y)]_{a,b}=
\left(\vartheta(a)\vartheta(b)\right)^{1/2}\int_x^y \overline{\phi}_a(z,t)\phi_b(z,t)\, dz$ and  $[\textsf{R}_T(x,y)]_{a,b}=\left(\vartheta(a)\vartheta(b)
\right)^{1/2}\overline{\phi}_a(x,t)\phi_b(y,t)$.

The second observation is that  (\ref{corrsillt}) can also be  generalized in the case of impenetrable anyons with $\kappa$ components  \cite{Patu19} 
that satisfy  the generalized commutation relations $\Psi_\sigma(x)\Psi_{\sigma'}^\dagger(y)+e^{-i\pi\varphi\mbox{\small{sgn}}(x-y)}
\Psi_{\sigma'}^\dagger(y) \Psi_\sigma(x)=\delta_{\sigma,\sigma'}\delta(x-y)$ and $\Psi_\sigma^\dagger(x)\Psi_{\sigma'}^\dagger(y)+e^{i\pi\varphi
\mbox{\small{sgn}}(x-y)}\Psi_{\sigma'}^\dagger(y)\Psi_\sigma^\dagger(x)=0$. Here $\varphi\in[0,1]$ is the statistics parameter and $\mbox{sgn}(x)=|x|/x$ 
with $\mbox{sgn}(0)=0$. The only modification that needs to  be made is to replace $\epsilon$ in (\ref{corrsillt}) with $-e^{-i\pi \varphi}$ for $x\le y$.  
The fermionic (bosonic) result is reproduced for $\varphi=0$ ($\varphi=1$).

The third observation is that while not completely obvious the representation (\ref{corrsillt}) also describes single component systems. For $\kappa=1$ 
the function $f(\sigma)=1$  and  (\ref{corrsillt}) contains as particular cases the Pezer-Buljan result \cite{PB07} for single component bosons at zero 
temperature and  the finite temperature  \cite{AGBK17} and anyonic \cite{Patu20} generalizations. 

Let us look at certain particular cases of (\ref{corrsillt}).

\textit{No magnetic field.} In this case all the chemical potentials are equal to $\mu$ and the Fermi function is $\vartheta(a)=(1+(1/\kappa) 
e^{(\varepsilon(a)-\mu)/T})^{-1}$. All the correlators are equal and given by 
\begin{align}\label{corrsilltb0}
\rho_\sigma(x,y)=&\det\left[\boldsymbol{1}-\left(1+\frac{\epsilon}{\kappa}\right)\, \textsf{V}_T+\frac{1}{\kappa} \textsf{R}_T\right]\nonumber\\
&\qquad\qquad -\det\left[\boldsymbol{1}-\left(1+\frac{\epsilon}{\kappa}\right)\, \textsf{V}_T\right]\, ,
\end{align}
with $\textsf{V}_T$ and $\textsf{R}_T$ given by (\ref{defvt}) and (\ref{defrt}).

\begin{widetext}
\textit{Zero temperature case: different chemical potentials.} Without loss of generality we consider the case of $\mu_1>\mu_2,\cdots,\mu_\kappa$.
At low temperatures the Fermi function becomes $\lim_{T\rightarrow 0}\vartheta(a)\sim \left(1+e^{(\varepsilon(a)-\mu_1)/T}\right)^{-1}$ which selects only 
the states  with $\varepsilon(a)\le\mu_1$. We consider  the number of states that satisfy this condition to be $N$. For the correlator of particles 
$\sigma=1$ we find 
\begin{align}
\rho_1(x,y)=&\det_N\left[\delta_{a,b}-(1+\epsilon)\int_x^y \overline{\phi}_a(z)\phi_b(z)\, dz
+\overline{\phi}_a(x)\phi_b(y)\right]
-\det_N\left[\delta_{a,b}-(1+\epsilon)\int_x^y \overline{\phi}_a(z)\phi_b(z)\, dz)\right]\, ,
\end{align}
which is the result derived by Pezer and Buljan \cite{PB07} for single component bosons in the case $\epsilon=1$. All the other correlators are zero because  
$\lim_{T\rightarrow 0}f(\sigma)=0$ for $\sigma\ne 1$. This shows that in the zero temperature limit a SILL system at different chemical potentials becomes 
fully polarized.

\textit{Zero temperature case: equal chemical potentials.} The Fermi function selects only the $N$ levels satisfying $\varepsilon(a)\le\mu$. All the 
correlators are equal to 
\be\label{corr0}
\rho_\sigma(x,y)=\det_N\left[\delta_{a,b}-\left(1+\frac{\epsilon}{\kappa}\right)\int_x^y \overline{\phi}_a(z)\phi_b(z)\, dz+\frac{1}{\kappa}
\overline{\phi}_a(x)\phi_b(y)\right]
-\det_N\left[\delta_{a,b}-\left(1+\frac{\epsilon}{\kappa}\right)\int_x^y \overline{\phi}_a(z)\phi_b(z)\, dz\right]\, .
\ee
\end{widetext}

\subsection{Dynamical fermionization of spinor gases}

\begin{figure*}
\includegraphics[width=\linewidth]{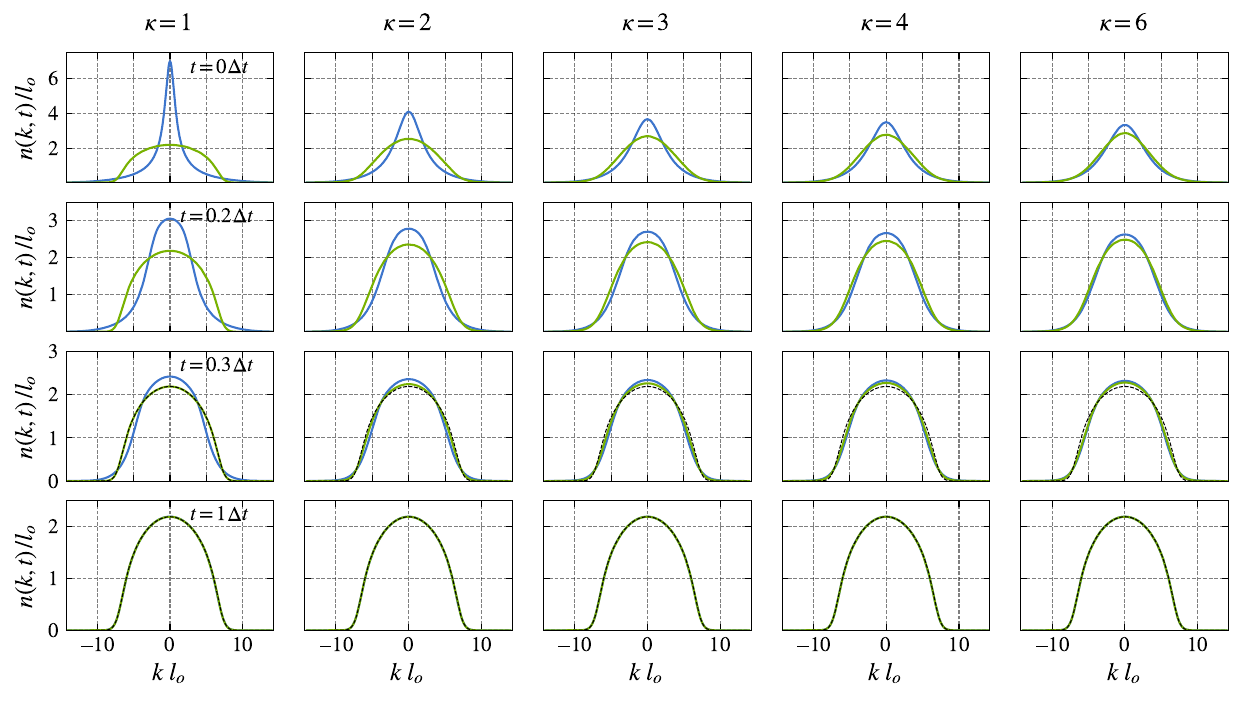}
\caption{Time evolution of the momentum distribution  for balanced systems of bosons (blue continuous line) and fermions (green continuous line) 
with $\kappa=\{1,2,3,4,6\}$ after release from a  harmonic trap with $m=\omega_0=1$ at finite temperature. Here $N=24$, $T=3\omega_0$ and the first row 
presents results for $t=0$, the second row for  $t=0.2\Delta t$, the third row for $t=0.3\Delta t$ and the fourth row for $t=0.2\Delta t$ with 
$\Delta t=2\pi/\omega_0$. In the third  and fourth rows the dashed black lines represent $n_{FF}^{\mu',T}(k)/\kappa$ with $n_{FF}^{\mu',T}(k)$ the 
momentum  distribution of trapped  spinless fermions at the same temperature and renormalized chemical potential given by Eq.~(\ref{renormalizedmu}). 
}
\label{fig5}
\end{figure*}

The determinant representations (\ref{corrsillt})  can be used to study the dynamics of a spinor gas after its release from a harmonic trap. This is 
a common  nonequilibrium scenario and in the case of the bosonic Tonks-Girardeau gas it was discovered that the asymptotic momentum distribution is 
equal to that of a system of free fermions in the initial harmonic trap. This phenomenon, called dynamical fermionization, was theoretically predicted 
in \cite{RM05,MG05} and experimentally confirmed in  \cite{WMLZ20}. In the case of spinor gases an analytical proof of dynamical fermionization at 
zero temperature can be found in \cite{ASYP21}. At finite temperature the  situation is more complex. In \cite{Patu23b} it was shown that the 
asymptotic momentum distribution of an impenetrable  spinor gas  with $\kappa$ components  at finite $T$ and equal chemical potentials, $\mu_\sigma=
\mu$,  approaches that of a system of fermions  $n_{FF}^{\mu',T}(k)=\sum_j|\phi_j(k)|^2/(1+e^{-[(j+1/2)-\mu']/T}) $ at the same  temperature but with 
a renormalized chemical potential 
\be\label{renormalizedmu}
\mu'=\mu+ T \ln \kappa\, .
\ee
Below, we will numerically verify this analytical prediction. This nonequilibrium scenario can be understood as a limiting case of a harmonic potential with 
time dependent frequency $V(z,t)=\omega^2(t) z^2/2$ with $\omega(t<0)=\omega_0$ and $\omega(t\ge 0)=0$. We consider an impenetrable spinor gas with 
$\kappa$ components, bosonic or fermionic, which is initially in thermal equilibrium  described by the grandcanonical ensemble at temperature $T$  
and  equal chemical potentials $\mu_\sigma=\mu,\  \sigma=1,\cdots,\kappa$. At $t<0$ the single particle orbitals are the Hermite functions $\phi_j(z)$ of 
frequency $\omega_0$. The time-evolution of the orbitals  is given by the scaling transformation (\cite{PP70}, Chap.VII of \cite{PZ98})
\begin{align}\label{scaling}
\phi_j(z,t)=\frac{1}{\sqrt{b(t)}}\phi_j\left(\frac{z}{b(t)},0\right)e^{i\frac{ x^2}{2}\frac{\dot{b}(t)}{b(t)}-i E_j\tau(t)}\, ,
\end{align}
with $b(t)$ the solution of the second-order differential equation $\ddot{b}+\omega^2(t) b=\omega_0^2/b^3$, also known as the  Ermakov-Pinney equation,  
with initial conditions $b(0)=1, \,  \dot{b}(0)=0$ and $\tau(t)=\int_0^t dt'/b^2(t')$. In our case the solution of the Ermakov-Pinney equation is $b(t)=
\left(1+\omega_0^2 t^2\right)^{1/2}$.  Inserting the scaling transformation (\ref{scaling}) in the expressions for the wavefunctions  (\ref{wavef}) then the 
formula for the correlator  (\ref{corrwavef}) becomes
\be\label{scalingr}
\rho_\sigma(x,y|\, t)=\frac{1}{b}\rho_\sigma\left(\frac{x}{b},\frac{y}{b}\left.\right|\, 0\right)
e^{-i\frac{\dot{b}}{b}\frac{m(x^2-y^2)}{2}}\, ,
\ee
from which the momentum distribution can be computed. In Fig.~\ref{fig5} we present the dynamics of the  momentum distribution for balanced bosonic  and 
fermionic systems after release from the trap. We consider systems with $\kappa=\{1,2,3,4,6\}$ and $N=24$ particles with $\omega_0=1$ and temperature $T=3\omega_0$. 
In both fermionic and bosonic cases after sufficient time the asymptotic momentum distribution becomes indistinguishable from the one for spinless free fermions 
at the same temperature and renormalized chemical potential defined in (\ref{renormalizedmu}).

\section{Determinant representations for the homogeneous SILL correlators}\label{s6}

\begin{figure*}[t]
\includegraphics[width=1\linewidth]{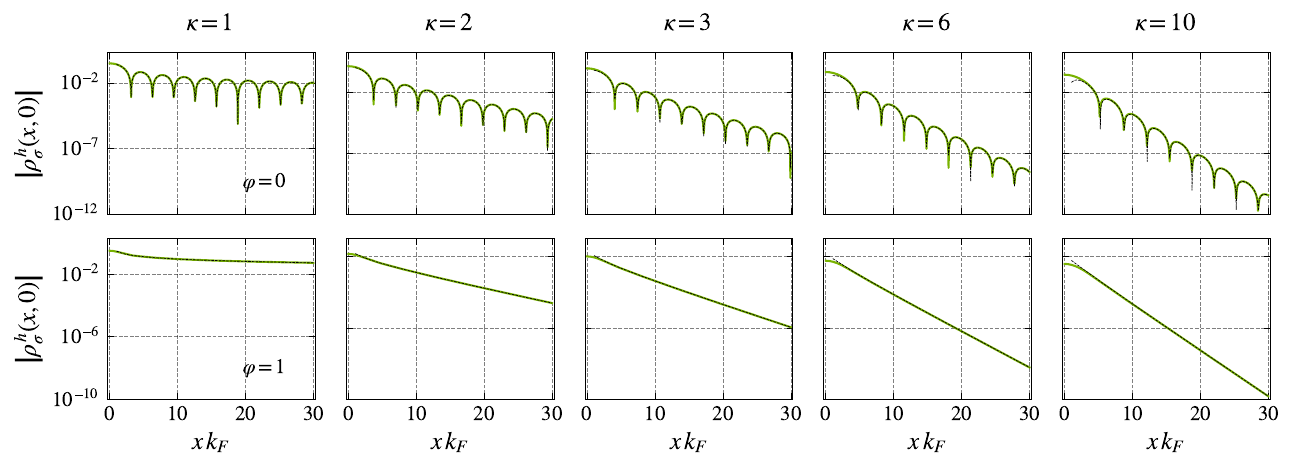}
\caption{Plots of the absolute value of the correlator $|\rho_\sigma^h(x,0)|$ (green continuous line) computed from the Fredholm determinant 
representation (\ref{corrsillh}) and the absolute value of the asymptotics (black dashed line) given by Eq.~(\ref{asympt}) for fermionic (first row) 
and bosonic  (second row) systems with $\kappa=\{1,2,3,6,10\}$ and $k_F=\pi D=1$.
}
\label{fig6}
\end{figure*}

The  representations (\ref{corrsillt}) derived in the previous section are valid for general trapping potentials or for systems with 
Dirichlet or Neumann boundary conditions. In the homogeneous case with periodic boundary conditions the Hamiltonian (\ref{ham}) is integrable and in principle one can apply 
the full power of Bethe ansatz techniques \cite{KBI} to derive similar results. While this should be in principle doable in this section we will show 
that by replacing  in (\ref{corrsillt}) the single particle orbitals of the  trapped system with the ones for the homogeneous system  we obtain the 
previously known representations for single  and two-component systems which lends credence to the argument that in fact this result is true for all 
values of $\kappa$.    

The single particle orbitals for the free fermionic system with periodic boundary conditions on a ring of circumference $L$ are  $\phi_a(z)=e^{i k_a z}/
\sqrt{L}$ with $k_a=2\pi a/L\, , a=0,\pm 1, \cdots$. We consider the case of zero temperature with equal chemical potentials  described by (\ref{corr0}) 
and arbitrary statistics  where $\epsilon=-e^{i\pi \varphi}$ with $\varphi\in [0,1]$. The fermionic (bosonic) case is recovered for $\varphi=0$ 
($\varphi=1$). We have
\begin{align}
\frac{1}{L}\int_x^ye^{i (k_b-k_a) z}\, dz&= \frac{2}{L}\frac{\sin\left[(k_b-k_a)(y-x)/2\right]}{k_b-k_a}\nonumber\\
&\ \ \times e^{i(k_b-k_a)y/2}e^{i(k_b-k_a)y/2}\, ,\\
\frac{1}{L}e^{-i k_a x}e^{i k_b y} &=\frac{1}{L}e^{i(k_a+k_b)(y-x)/2}\nonumber\\
&\ \ \times e^{i(k_b-k_a)y/2}e^{i(k_b-k_a)y/2}\, ,
\end{align}
where the $e^{i(k_b-k_a)y/2}e^{i(k_b-k_a)y/2}$ factors can be discarded because they are just a similarity transformation. Taking the thermodynamic
limit  $N,L\rightarrow\infty$ such that $D=N/L$ we find ($x\le y$)
\begin{align}\label{corrsillh}
\rho_\sigma^h(x,y)=&\det\left[\boldsymbol{1}-\left(1-\frac{e^{i\pi \varphi}}{\kappa}\right)\, \hat{\textsf{v}}+\frac{1}{\kappa}\hat{\textsf{r}}\right]\nonumber\\
&\qquad\qquad -\det\left[\boldsymbol{1}-\left(1-\frac{e^{i\pi \varphi}}{\kappa}\right)\, \hat{\textsf{v}}\right]\, ,
\end{align}
with the result expressed in terms of Fredholm determinants of the integral operators $\hat{\textsf{v}}$ and $\hat{\textsf{r}}$ acting on $[-k_F,k_F]$
($k_F=\pi D$) with kernels
\begin{align}
\textsf{v}(k,k')&=\frac{\sin\left[(k-k')(y-x)/2\right]}{\pi (k-k')}\, ,\\
\textsf{r}(k,k')&=\frac{e^{i(k+k')(y-x)/2}}{2\pi}\, .
\end{align}
The action  of $\hat{\textsf{v}}$, and similarly of $\hat{\textsf{r}}$,  on an arbitrary function $\phi(k)$ is given by $(\hat{\textsf{v}} \phi)(k)=
\int_{-k_F}^{k_F}\textsf{v}(k,k')\phi(k')\, dk'$. In the single component case, $\kappa=1$, the representation (\ref{corrsillh}) is equivalent with 
the result for impenetrable bosons derived by Schultz \cite{Schu63} and Lenard \cite{L66}, at $\varphi=0$  is equal to $\sin[k_F(y-x)]/\pi(y- x)$, which 
is the well known correlator for free fermions, and for arbitrary $\varphi$  is the same as the result derived in \cite{PKA08} for impenetrable anyons. 
For the 
two-component system, $\kappa=2$, (\ref{corrsillh}) agrees with the result derived by Izergin and Pronko \cite{IP98} for fermionic and bosonic spinor
gases and the result derived in \cite{Patu19} for  anyonic two-component gases (note that the result obtained in this section was derived under the 
assumption that $x\le y$, for $x>y$ one should take the complex conjugate of (\ref{corrsillh})). Therefore, we conjecture that the determinant representation 
(\ref{corrsillh}) is also valid when $\kappa>2$.

The large distance asymptotics of the static correlators can be rigorously computed using a method similar to the one employed in \cite{CZ2,Patu19}
for two-component systems. The main ingredient is a very powerful result regarding the asymptotics of the generalized sine-kernel derived by  Kitanine,  
Kozlowski,  Maillet,  Slavnov and Terras in \cite{KKMST09}. Introducing two parameters
\be
\xi=-\left(1-\frac{e^{i\pi \varphi}}{\kappa}\right)\, ,\ \ \ \nu=-\frac{1}{2\pi i}\ln(1+\xi)=-i\frac{\ln\kappa}{2\pi}-\frac{\varphi}{2}\, ,
\ee
the asymptotics are ($x\rightarrow\infty$)
\begin{align}\label{asympt}
\rho_\sigma^h(x,0)&=\frac{1}{\kappa}\frac{\pi e^{\mathcal{C}(\nu)}}{\xi \sin(\pi \nu)}\frac{e^{-2i k_F\nu x}}{x^{2\nu^2+1}}\nonumber\\
& \times \left[\frac{(2k_F x)^{-2\nu}}{\Gamma^2(-\nu)}e^{-i k_F x}\right.
 \left.-\frac{(2k_F x)^{2\nu}}{\Gamma^2(\nu)}e^{i k_F x}\right]\, ,
\end{align}
with
$
\mathcal{C}(\nu)=-2\nu^2\left[1+\ln(2k_F)\right]+2\nu \ln\left[\frac{\Gamma(\nu)}{\Gamma(-\nu)}\right]
-2\int_0^\nu \ln\left[\frac{\Gamma(t)}{\Gamma(-t)}\right]\, dt\, , 
$
and  $\Gamma(t)$ is the Gamma function. For $\kappa=1$ the asymptotics Eq.~(\ref{asympt}) reproduce the  known results for impenetrable 
single component bosons \cite{VT79,JMMS80, Gang04},  impenetrable anyons \cite{CM07,Patu19} and by taking the limit $\varphi\rightarrow 0$ one obtains  
$\sin[k_F x]/\pi x$ which is the exact result  for free fermions. For two-component systems, $\kappa=2$, one obtains the result for fermions derived in 
\cite{BL,CZ2,FB04}  for bosons derived in \cite{CSZ05,Patu19}  and for impenetrable anyons in \cite{Patu19}.
For  bosonic and anyonic systems described by a statistics parameter $\varphi\sim 1$ the first term in the right hand side of (\ref{asympt}) gives the 
main  contribution but in the case of fermionic systems and for anyonic systems with $\varphi\sim 0$ both terms of the expansion are important. It is 
important to note that our result also contains the constants in front of each term. In general these constants cannot be obtained using LL/bosonization  
\cite{H81,G03} or the first quantized path-integral representation for the correlators \cite{FB04,Fiet07}. The main feature of the asymptotics is the 
zero temperature exponential decay $e^{-x k_F \ln \kappa /\pi}$  with an exponent proportional to the logarithm of the number of components of the system 
\cite{Fiet07}. 
In Fig.~\ref{fig6} we present results for bosonic and fermionic systems with different number of components evaluated numerically \cite{Born10} using the 
Fredholm  determinant representation (\ref{corrsillh}) compared with the predictions of the asymptotic formula (\ref{asympt}). We see that we have excellent 
agreement, especially  in the bosonic case, even though we have considered only the first two terms of the expansion.

In the case of homogeneous spin-$1/2$ fermions, Matveev, Furusaki, and Glazman \cite{MFG07a, MFG07b} introduced 
an elegant method for computing the spectral function in both the Luttinger liquid  and spin-incoherent regimes. 
Their approach involves bosonizing the charge degrees of freedom while treating the spin excitations exactly. 
Notably, they observed that, in addition to the expected features at $k_F$ and $3k_F$, the spectral function exhibits 
a pronounced peak centered at $k = 0$ (see also \cite{FQTH05}). Whether this feature persists in systems with more 
than two components remains an open question. Addressing this would require generalizing the methods presented in 
this paper to dynamic correlators in such systems.

\section{Conclusions}\label{s7}

In this paper we have introduced an efficient  method of computing the  multidimensional integrals that appear in the expressions for  the 
correlation functions of the strongly interacting spinor gases in 1D. While we have focused on the case of fermionic and bosonic  gases it should be 
mentioned that our results can be applied in the case of Bose-Fermi  mixtures \cite{ID06,DJAR17,DBBR17} which present similar spin-charge factorizations 
of the correlators with the charge functions having identical definitions. Using this method we were able to investigate systems with a larger number 
of particles than considered before and we have shown that small changes in temperature have dramatic effects on the static and dynamic properties of 
strongly interacting spinor gases. We derived determinant representations for the correlation functions of trapped and  homogeneous systems 
in the spin-incoherent regime and in the latter case we determined the large distance asymptotics. We expect that our results to be  generalizable 
in the case of the density-density correlation functions and the  calculation of the full counting statistics. This is deferred to a future publication.

\acknowledgments

Financial support from the Grant No. 30N/2023 of the National Core Program of the Romanian Ministry of Research, Innovation and Digitization is gratefully 
acknowledged.

\newpage

\appendix

\begin{widetext}

\section{Fourier integral expression for the single particle densities}\label{app1}

In this Appendix we will rewrite the expression for the single particle densities in a Fourier integral form which can be evaluated numerically in 
a simple fashion. The first  observation that we make is that due to the fact that the product $\overline{\psi}_F\psi_F$ appearing in  
Eq.~(\ref{singled}) vanishes when two coordinates are equal and is symmetric independently  in $z_1,\cdots, z_{d-1}$ and $z_{d+1},\cdots,z_{N}$  
we can extend the  integration to  $\tilde{\Gamma}_d=L_-\le z_1,\cdots, z_{d}<x<z_{d+1},\cdots,z_N\le L_+ $ by multiplying  the integral with  
$1/(d-1)!(N-d)!$ obtaining
\begin{align}
\rho_d(x)&=\frac{1}{(d-1)!(N-d)!}\int_{\tilde{\Gamma}_{d}} \prod_{\substack{j=1\\ j\ne d}}^N  dz_j
\sum_{P\in S_N}\sum_{P'\in S_N} (-1)^{P+P'}
\left(\prod_{k=1}^{d-1}\overline{\phi}_{P_k}(z_k)\phi_{{P'}_k}(z_k)\right) \overline{\phi}_{P_d}(x)\phi_{{P'}_d}(x)\nonumber\\
&\qquad\qquad\qquad\qquad\qquad\qquad\qquad\times\left(\prod_{m=d+1}^{N}\overline{\phi}_{P_m}(z_m)\phi_{{P'}_m}(z_m)\right)\, . 
\end{align}
Introducing three $N\times N$ matrices defined in (\ref{defm0}), (\ref{defm1}) and  (\ref{defmr}) and using the fact that every  permutation  $P'$  
can be written as $P'=QP$ with $Q$ another permutation the previous relation can be rewritten 
as
\begin{align}\label{a10}
\rho_d(x)&=\frac{1}{(d-1)!(N-d)!}
\sum_{P\in S_N}\sum_{Q\in S_N} (-1)^{Q}
\left(\prod_{k=1}^{d-1}\textsf{M}^0_{P_k,QP_k}\right)\textsf{M}^r_{P_d,QP_d}\left(\prod_{m=d+1}^{N}\textsf{M}^1_{P_m,QP_m}\right)\, . 
\end{align}
We will show that (\ref{a10}) is equivalent to 
\begin{align}\label{a11}
\rho_d(x)&=\int_{0}^{2\pi}\frac{d\alpha}{2\pi}e^{-i(d-1)\alpha}\int_{0}^{2\pi}\frac{d\beta}{2\pi}e^{-i\beta}
\sum_{Q\in S_N}(-1)^N\prod_{k=1}^N\left(e^{i\alpha} \textsf{M}^0_{k,Q_k}+e^{i\beta}\textsf{M}^r_{k,Q_k}+\textsf{M}^1_{k,Q_k}\right)\, ,
\end{align}
expression which contains two auxiliary phase variables \cite{IP98,ID06,Patu22}. 
We define a set of equivalence classes on the set of permutations of $N$ elements as follows. For a given $d$ two permutations $R$  and $R'$ are equivalent, 
denoted by $R\sim R'$, if $R=(R_1,\cdots,R_{d-1},R_d,R_{d+1},\cdots,R_N)$, $R'=(R'_1,\cdots,R'_{d-1}, R'_d,R'_{d+1},\cdots,R'_N)$ and $\{R_1,\cdots,R_{d-1}
\}=\{R'_1,\cdots,R'_{d-1}\}$ and $\{R_{d+1},\cdots, R_N\}=\{R'_{d+1},\cdots,  R'_N\}$. This implies that $R_d=R'_d$. For example, if $N=6$ and $d=4$, 
$R=(521346)$ and $R'=(215364)$ are equivalent. We will call  the representative element of a class of equivalence the permutation in which $\{R_1,\cdots,R_{d-1}
\}$ and $\{R_{d+1},\cdots, R_N\}$ are ordered.  For the previous example the representative element is $\tilde{R}=(125346).$ 
For two  equivalent permutations, in the sense defined above, the sum $\sum_{Q\in S_N}(-1)^Q\left[\cdots\right]$ in (\ref{a10}) gives the same  result. Each 
class of equivalence has $(d-1)!(N-d)!$ elements which means that (\ref{a10}) can be written as 
\begin{align}\label{a12}
\rho_d(x)&=
\sum_{\tilde{P}}\sum_{Q\in S_N} (-1)^{Q}
\left(\prod_{k=1}^{d-1}\textsf{M}^0_{\tilde{P}_k,Q\tilde{P}_k}\right)\textsf{M}^r_{\tilde{P}_d,Q\tilde{P}_d}\left(\prod_{m=d+1}^{N}
\textsf{M}^1_{\tilde{P}_m,Q\tilde{P}_m}\right)\, , 
\end{align}
where the first sum is over the representative elements which have cardinality $N!/[(d-1)(N-d)!]$. In  (\ref{a11}) each term 
selected by the integration over $\alpha$ and $\beta$ is in one-to-one correspondence with the equivalence  classes of permutations (note that the integrations 
select $(d-1)$ terms of $\textsf{M}^0$ one  of $\textsf{M}^r$ and $(N-d)$ of  $\textsf{M}^1$). For example, considering $N=6$, $d=4$ and $\tilde{P}=(125346)$  the 
one-to-one term is  $\left(e^{3i\alpha}\textsf{M}^0_{1,Q_1}\textsf{M}^0_{2,Q_2}\textsf{M}^0_{5,Q_5}\right)\left(e^{i\beta}\textsf{M}^r_{3,Q_3}\right) \left
(\textsf{M}^1_{4,Q_4}\textsf{M}^1_{6,Q_6}\right)$. This shows that (\ref{a10}) and (\ref{a11}) are the same. The integrand in (\ref{a11}) can be written as a 
determinant obtaining the final expression (\ref{a13}).

\section{Fourier integral expression for the local exchange coefficients }\label{app2}

The derivation of the Fourier integral expression for the local exchange coefficients (\ref{coeff}) starts by  integrating over the delta function 
obtaining
\be\label{a15}
J_d^0=N!\int_{\Gamma_d} \prod_{\substack{j=1 \\j\ne d}}^N dz_j\left|\frac{\6 \psi_F(\boldsymbol{q}^0)}{\6 z_d }\right|^2_{z_d=z_{d+1}}\, , 
\ee
with $\Gamma_d=L_-\le z_1<\cdots<z_{d-1}<z_{d+1}<\cdots<z_N\le L_+ $. The integrand is symmetric in $z_1,\cdots,z_{d-1}$ and $z_{d+1},
\cdots,z_N$, so we can extend the domain of integration to $\tilde{\Gamma}_d=L_-\le z_1,\cdots,z_{d-1}<z_{d+1},\cdots,z_N
\le L_+ $  after multiplication with  $1/[(d-1)!(N-d-1)!]$. We find
\begin{align}\label{a15b}
J_d^0&=\frac{N!}{(d-1)!(N-d-1)!}\int_{L_-}^{L_+}dz_{d+1}\int_{L_-}^{z_{d+1}}\prod_{k=1}^{d-1} dz_k\int_{z_{d+1}}^{L_+}\prod_{m=d+2}^N 
dz_m
\left|\frac{\6 \psi_F}{\6 z_d }\right|^2_{z_d=z_{d+1}} \, ,\nonumber\\
&=\frac{1}{(d-1)!(N-d-1)!}\int_{L_-}^{L_+}d\xi \sum_{P\in S_N}\sum_{P'\in S_N}(-1)^{P+P'}
\left(\int_{L_-}^\xi \prod_{k=1}^{d-1}dz_k\overline{\phi}_{P_k}(z_k)\phi_{P'_k}(z_k)\right)\left(\overline{\phi}'_{P_d}(\xi)\phi'_{P'_d}
(\xi)\right)\nonumber\\
&\qquad\qquad\qquad\qquad\qquad\qquad\qquad\qquad\times \left(\overline{\phi}_{P_{d+1}}(\xi)\phi_{P'_{d+1}}(\xi)\right) 
\left(\int_{\xi}^{L_+}  \prod_{m=d+2}^{N}dz_m\overline{\phi}_{P_m}(z_m)\phi_{P'_m}(z_m)\right)\, ,
\end{align}
where in the second line we introduced $z_{d+1}=\xi$. Writing $P'=QP$  the previous expression can be written in terms of elements of the 
matrices $\textsf{M}^{0,1,r,d}(\xi)$ defined in (\ref{defm0}), (\ref{defm1}), (\ref{defmr})  and (\ref{defmdj}) as
\begin{align}\label{a16}
J_d^0&=\int_{L_-}^{L_+}d\xi\, \frac{1}{(d-1)!(N-d-1)!}\sum_{P\in S_N}\sum_{Q\in S_N}(-1)^{Q}
\left(\prod_{k=1}^{d-1}\textsf{M}^0_{P_k,QP_k}(\xi)\right)\textsf{M}^d_{P_d,QP_d}(\xi)\textsf{M}^r_{P_{d+1},QP_{d+1}}(\xi)\nonumber\\
&\qquad\qquad\qquad\qquad\qquad\qquad\qquad\qquad\qquad\qquad\qquad\qquad\qquad\qquad\times \left(\prod_{m=d+2}^{N}\textsf{M}^1_{P_m,QP_m}
(\xi)\right)\, .
\end{align}
We will denote the integrand appearing in (\ref{a16}) by $I_d(\xi)$. Similar to the case of the single particle densities an equivalent 
expression  of $I_d(\xi)$ can be derived introducing three auxiliary phases with the result
\begin{align}
I_d(\xi)&=\int_{0}^{2\pi}\frac{d\alpha}{2\pi}e^{-i(d-1)\alpha} \int_{0}^{2\pi}\frac{d\beta}{2\pi}e^{-i\beta} \int_{0}^{2\pi}\frac{d\gamma}
{2\pi}e^{-i\gamma}\nonumber\\
&\qquad\qquad\qquad\qquad\times\sum_{Q\in S_N}(-1)^Q\prod_{k=1}^N\left(e^{i\alpha}\textsf{M}^0_{k,Q_k}(\xi)+
e^{i\beta}\textsf{M}^r_{k,Q_k}(\xi)+e^{i\gamma}\textsf{M}^d_{k,Q_k}(\xi) +\textsf{M}^1_{k,Q_k}(\xi)\right)\, ,
\end{align}
which is exactly Eq.~(\ref{c1}).

\section{Fourier integral expression for the one-body density matrix elements}\label{app3}

The one-body density matrix elements are defined in  (\ref{obdm}). We consider the case $x\le y$  and $d_1\le d_2$ . The integrand is symmetric in three set of 
variables $z_1,\cdots,z_{d_1-1}$;    $z_{d_1+1},\cdots,z_{d_2}$ and $z_{d_2+1},\cdots,z_N$ and is zero when two of them are equal. Therefore, we can extend the  
domain of integration to \[\tilde{\Gamma}_{d_1,d_2}(x,y)={L_-\le z_1,\cdots, z_{d_1-1}<x<z_{d_1+1},\cdots,z_{d_2}<y \\ <z_{d_2+1},\cdots,z_N\le L_+}\, ,\]
by multiplying the integral with $1/[(d_1-1)!(d_2-d_1)!(N-d_2)!]$. We obtain 
\begin{align}\label{a20}
\rho_{d_1,d_2}(x,y)&=\frac{1}{(d_1-1)!(d_2-d_1)!(N-d_2)!}\int_{\tilde{\Gamma}_{d_1,d_2}(x,y)}\prod_{\substack{j=1\\ j\ne d_1}}^N dz_j
\sum_{P\in S_N}\sum_{P'\in S_N} (-1)^{P+P'} \left(\prod_{k=1}^{d_1-1}\overline{\phi}_{P_k}(z_k)\phi_{P'_k}(z_k)\right)\nonumber\\
&\qquad\qquad\qquad\qquad\times  \left(\overline{\phi}_{P_d}(x)\phi_{P'_d}(y)\right)\left(\prod_{m=d_1+1}^{d_2}\overline{\phi}_{P_m}(z_m)\phi_{P'_m}(z_m)\right)
\left(\prod_{n=d_2+1}^{N}\overline{\phi}_{P_n}(z_n)\phi_{P'_n}(z_n)\right)\, .
\end{align}
Writing $P'=QP$ the right hand side of  (\ref{a20}) becomes 
\begin{align}\label{a21}
\rho_{d_1,d_2}(x,y)&=\frac{1}{(d_1-1)!(d_2-d_1)!(N-d_2)!}
\sum_{P\in S_N}\sum_{Q\in S_N} (-1)^{Q} \left(\prod_{k=1}^{d_1-1}\textsf{M}^0_{P_k,QP_k}(x)\right)\textsf{M}^n_{P_{d_1},QP_{d_1}}(x,y)\nonumber\\
&\qquad\qquad\qquad\qquad\times  \left(\prod_{m=d_1+1}^{d_2}\textsf{M}^2_{P_m,QP_m}(x,y)\right)\left(\prod_{n=d_2+1}^{N}\textsf{M}^1_{P_n,QP_n}(y)\right)\, ,
\end{align}
with the $\textsf{M}^{0,1,2,n}$ matrices defined in (\ref{defm0}), (\ref{defm1}), (\ref{defm2}) and (\ref{defmn}).
Introducing three phases this last identity  can be shown to be equivalent to 
\begin{align}\label{a21b}
\rho_{d_1,d_2}(x,y)&=\int_{0}^{2\pi}\frac{d\alpha}{2\pi}e^{-i(d_1-1)\alpha} \int_{0}^{2\pi}\frac{d\gamma}{2\pi}e^{-i\gamma}
\int_{0}^{2\pi}\frac{d\beta}{2\pi}e^{-i(d_2-d_1)\beta}\nonumber\\
&\qquad\qquad\qquad\qquad\sum_{Q\in S_N}(-1)^Q\prod_{k=1}^N\left(e^{i\alpha} \textsf{M}^0_{k,Q_k}+e^{i\gamma} \textsf{M}^n_{k,Q_k}+e^{i\beta} \textsf{M}^2_{k,Q_k}
+ \textsf{M}^1_{k,Q_k}\right)(x,y)\, ,
\end{align}
which is Eq.~(\ref{a22}) of the main text.

\section{Partition function for the impenetrable spinor gases}\label{app4}

The partition function of 1D impenetrable spinor gases is independent of statistics. It is instructive to consider first the particular case $\kappa=3$. As we will 
see the generalization for arbitrary $\kappa$ follows easily from this particular example. The partition function in the grandcanonical ensemble for $\kappa=3$ is
\begin{align}\label{ap1}
Z_{\kappa=3}&=\sum_{N=0}^\infty \sum_{q_1<\cdots< q_N} \sum_{N_1=0}^{N}\sum_{N_2=0}^{N-N_1} 
\sum_{l=1}^{N!/[N_1!N_2!N_3!]} e^{-\sum_{j=1}^N\varepsilon(q_j)/T+\sum_{j=1}^3\mu_j N_j/T}\, .
\end{align}
Because $N!/[N_1!N_2!N_3!]=C^N_{N_1}C^{N-N_1}_{N_2}$ ($N=N_1+N_2+N_3$) the sum over the spin eigenstates can be written as
\begin{align}
\sum_{N_1=0}^{N}\sum_{N_2=0}^{N-N_1} C^N_{N_1}C^{N-N_1}_{N_2} e^{\sum_{j=1}^3 \mu_j N_j/T}&=
\sum_{N_1=0}^{N}C^N_{N_1} e^{\mu_1 N_1/T } \left(e^{\mu_2/T}+e^{\mu_3/T}\right)^{N-N_1}\, ,\nonumber\\
&= \left(e^{\mu_1/T}+e^{\mu_2/T}+e^{\mu_3/T}\right)^{N}\, .
\end{align}
Plugging this result in (\ref{ap1}) we obtain 
\begin{align}\label{ap2}
Z_{\kappa=3}&=\sum_{N=0}^\infty \sum_{q_1<\cdots< q_N} \left(e^{\mu_1/T}+e^{\mu_2/T}+e^{\mu_3/T}\right)^{N}
  e^{-\sum_{j=1}^N\varepsilon(q_j)/T}\, ,\nonumber\\
&=\prod_{q=1}^\infty\left[1+\left(e^{\mu_1/T}+e^{\mu_2/T}+e^{\mu_3/T}\right)e^{-\varepsilon(q)/T}\right]\, .
\end{align}
The natural generalization for arbitrary $\kappa$ is 
\begin{align}\label{partition}
Z
&=\prod_{q=1}^\infty\left[1+\left(\sum_{\sigma=1}^\kappa e^{\frac{\mu_\sigma}{T}}\right)e^{-\varepsilon(q)/T}\right]\, .
\end{align}

\end{widetext}

\end{document}